\documentclass[prx,times, twocolumn, secnumarabic, superscriptaddress, longbibliography, balancelastpage, hyperref=pdftex, notitlepage,floatfix]{revtex4-1}

\usepackage{amsmath} % AMS Math Package
\usepackage{amsthm} % Theorem Formatting
\usepackage{amssymb}	% Math symbols such as \mathbb
\usepackage{bbm}
\usepackage{color}         % produces boxes or entire pages with colored backgrounds
\usepackage{graphics}      % standard graphics specifications
\usepackage[pdftex]{graphicx}      % alternative graphics specifications
\usepackage{epsf}          % old package handles encapsulated post script issues
\usepackage{bm}            % special 'bold-math' package
\usepackage{verbatim}			% for comment environment
\usepackage{multirow}
\usepackage{subfigure}
\usepackage[T1]{fontenc}
\usepackage[colorlinks=true]{hyperref} % this package should be added after all others

\newcommand{\avg}[1]{\left< #1 \right>} % for average
\newcommand{\ket}[1]{\left| #1 \right>} % for Dirac bras
\newcommand{\bra}[1]{\left< #1 \right|} % for Dirac kets
\newcommand{\braket}[2]{\left< #1 \vphantom{#2} \right|
 \left. #2 \vphantom{#1} \right>} % for Dirac brackets
\newcommand{\ketbra}[2]{\mbox{$\left| #1 \vphantom{#2} \right\rangle \!\! \left\langle \vphantom{#1} #2 \right|$}}
\let\baraccent=\= % rename builtin command \= to \baraccent
\renewcommand{\=}[1]{\stackrel{#1}{=}} % for putting numbers above =
\newcommand{\lsim}{\mathrel{\rlap{\lower4pt\hbox{\hskip1pt$\sim$}}
    \raise1pt\hbox{$<$}}}                % less than or approx. symbol
\newcommand{\gsim}{\mathrel{\rlap{\lower4pt\hbox{\hskip1pt$\sim$}}
    \raise1pt\hbox{$>$}}}                % greater than or approx. symbol

\newcommand{\eg}{e.g.}

\newcommand{\rmd}{{\rm d}}
\newcommand{\rmi}{{ i}}
\newcommand{\rme}{{\rm e}}
\newcommand{\tr}{{\rm tr}}
\newcommand{\dg}{\dagger}
\renewcommand{\Re}{{\rm Re}}

\theoremstyle{definition}

\theoremstyle{remark}

\newcommand{\s}{\mathcal{P}}
\newcommand{\e}{\mathcal{S}}
\newcommand{\se}{\mathcal{PS}}

\begin{document}

\title{Using a biased qubit to probe complex systems}
\author{Felix A. Pollock}
\affiliation{School of Physics \& Astronomy, Monash University, Victoria 3800, Australia}

\author{Agata Ch\k{e}ci\'nska}
\affiliation{Institute of Theoretical Physics, University of Warsaw, Pasteura 5, 02-093 Warsaw, Poland}

%\author{Alex Monr\'as}
%\affiliation{F\'isica Te\'orica: Informaci\'o i Fen\'omens Qu\'antics, Universitat Aut\'onoma de Barcelona, 08193 Bellaterra (Barcelona)}

\author{Saverio Pascazio}
\affiliation{Dipartimento di Fisica and MECENAS, Universit\`a di Bari, I-70126 Bari, Italy}
\affiliation{INFN, Sezione di Bari, I-70126 Bari, Italy}

\author{Kavan Modi}
\affiliation{School of Physics \& Astronomy, Monash University, Victoria 3800, Australia}
\date{\today}

\begin{abstract}
Complex mesoscopic systems play increasingly important roles in modern science -- from understanding biological functions at the molecular level, to designing solid-state information processing devices. The operation of these systems typically depends on their energetic structure, yet probing their energy-landscape can be extremely challenging; they have many degrees of freedom, which may be hard to isolate and measure independently. Here we show that a qubit (a two-level quantum system) with a biased energy-splitting can directly probe the spectral properties of a complex system, without knowledge of how they couple. Our work is based on the completely-positive and trace-preserving map formalism, which treats any unknown dynamics as a `black-box' process. This black box contains information about the system with which the probe interacts, which we access by measuring the survival probability of the initial state of the probe as function of the energy-splitting and the process time. Fourier transforming the results yields the energy spectrum of the complex system. %, and the results can further determine its initial coherence. 
Without making assumptions about the strength or form of its coupling, our probe could 
determine aspects of a complex molecule's energy landscape as well as, in many cases, test for coherent superposition of its energy eigenstates.
\end{abstract}

\maketitle

%%%%%%%%%%%%%%%%%%%%%%%%
%%%%%%%%%%%%%%%%%%%%%%%%
\section{Introduction}
%%%%%%%%%%%%%%%%%%%%%%%%
%%%%%%%%%%%%%%%%%%%%%%%%

Measuring the properties of complex systems at the mesoscopic level is extremely difficult and of great importance. By definition, these systems are too small to admit an effective statistical description, but too large for their dynamically relevant degrees of freedom to be straightforwardly isolated. This is especially true at the classical--quantum boundary, where quantum effects are still important, but understanding their dynamical and energetic consequences can become highly non-trivial due to the size of the system \cite{NitzanBook}. Such systems play a vital role in the function of living organisms \cite{fmoreview, vlatkobirds, arXiv:1409.4522, quantumnose,mitochondria,dna1,dna2}, and increasingly form the basis for quantum technologies \cite{nielsen}.  Over the last few decades, much effort has gone into developing experimental techniques to characterise and quantify the salient features of complex systems in a variety of contexts. For example, pump-probe and multidimensional spectroscopy are used to isolate electronic contributions to the energetic structure of complex molecules \cite{Mukamel, Scholes} and solid state systems \cite{Goulielmakis}, while techniques utilising Raman scattering are used to determine vibrational spectra \cite{RamanMolecules, RamanGraphene}. In the same spirit, there have been several recent attempts to quantify quantum coherence in complex systems \cite{Girolami2014, arXiv:1311.0275, arXiv:1412.7161, arXiv:1312.5724, Kassal}.

To learn anything about a system, it must first interact with a probe, which induces open dynamics for both the probe and the system. From the perspective of the probe, the complex system acts as a black box, which evolves the probe's quantum state in a generally non-unitary way. Information about the complex system must then be determined by studying this black-box dynamics. 
The mathematical tools that describe open quantum dynamics operationally have a rich history \cite{petruccione}. Foremost among them is the completely-positive trace-preserving (CPTP) dynamical map \cite{sudmaps, kraus}, which describes any transformation between two temporal points in full generality -- the map $\Lambda_{\tau:0}$ relates the probe's density operator at the end of the evolution, $\rho_\tau$, to that at the beginning, $\rho_0$. The map contains all the dynamical information that is experimentally accessible by making measurements on the probe only, and we can ask what one can infer from it about the Hamiltonian of the complex system $H_\e$ and its interaction with the probe $V$. Stinespring's dilation theorem tells us that any CPTP map stems from unitary evolution of the probe and system together \cite{petruccione}:
\begin{gather}\label{Stine}
\rho_\tau = \Lambda_{\tau:0}[\rho_0] = \tr_\e\left[ \rme^{-\rmi H_{\text{tot}}\, \tau} \rho_0 \otimes \rho_\e \, \rme^{\rmi H_{\text{tot}} \, \tau} \right],
\end{gather}
where $H_{\text{tot}} = H_\s + H_\e + V$, $H_\s$ can be identified as the Hamiltonian of the probe, $\rho_\e$ is the initial state of the system (which we will always assume to be uncorrelated from the probe initially), $\tau$ is the evolution time and we have set $\hbar=1$. However, this relationship is not unique; in fact, without additional assumptions about the nature of the system, there is little that can be directly deduced from $\Lambda_{\tau:0}$ -- the dynamics of the probe alone -- about $\rho_\e$ and $V_\se \equiv H_\e + V$.  

Here, we develop a new method for probing, in principle, \emph{any} 
uncharacterised complex system, be it a large organic molecule or some exotic meta-material, using the simplest quantum probe -- a qubit, or two-level quantum system. Several schemes already exist which use qubits to probe larger systems: impurities in Bose-Einstein condensates can be used to determine phase \cite{BECqdotBruderer}, phonon spectra \cite{BECimpurityHangleiter} and temperature \cite{BECthermometrySabin}; in cavity and circuit quantum electrodynamics, effective two-level systems have been proposed as probes for environmental noise spectra \cite{Mueller2015, Norris2015} and witnesses for strong coupling \cite{Felicetti2015}; the use of single spins to precisely map complex electric and magnetic fields \cite{Cole2009} and environmental coherence \cite{Jeske2012} has also been proposed. However, in each of these cases, the analysis relies on a particular form for the probe-system Hamiltonian $V_\se$. Our result differs, and is more generally applicable, in that %we assume that
it applies even when $V_\se$ is completely unknown, though in many situations it may be limited by an inability to control the probe sufficiently.  

It should be noted that $V_\se$ necessarily includes those degrees of freedom which would usually be considered as an environment to the system of interest. As long as these do not strongly couple directly to the probe, their effect will generally be to smear out the energy levels of $\e$.

\begin{figure}[t]
%\makebox[220pt]{
\includegraphics[width=220pt, viewport=0 0 220 250]%660pt 750pt]
{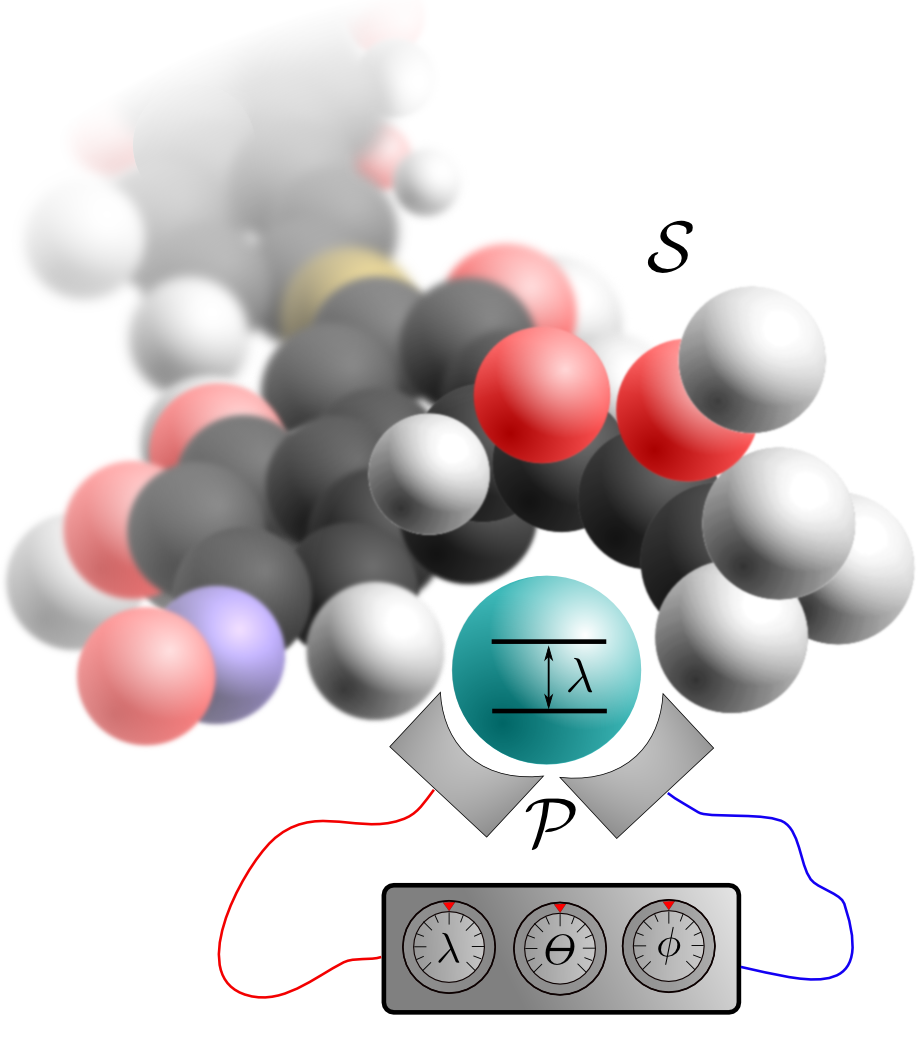}%}
\caption{(Color online) {\bf The controlled two-level probe scenario:}  The properties of an uncharacterised complex system $\e$ can be inferred from the way in which it influences dynamics of a two-level probe $\s$, with which it interacts. By varying a controllable bias of magnitude $\lambda$, more information about the system becomes accessible, including aspects of its state and spectrum.} \label{fig:sysenv}\vspace{-10pt}
\end{figure}

The scenario we consider in this article
is illustrated in Fig.~\ref{fig:sysenv}. We allow a classical control field with variable magnitude (denoted by $\lambda$) to modulate the probe's energy splitting in a basis of our choosing, parameterised by the angles $\theta$ and $\phi$, leading to a Hamiltonian for the probe 
\begin{gather}
H_\s = \frac{1}{2}\lambda %\,\epsilon 
\, \sigma_{(\theta,\phi)},
\end{gather}
where $\sigma_{(\theta,\phi)}= \sin\theta \cos\phi \, \sigma_x + \sin\theta \sin\phi \, \sigma_y +  \cos\theta \, \sigma_z$ is the Pauli operator that determines the basis in which the bias is applied%; $\epsilon$ sets the energy scale of $H_\s$
. Moreover, we are free to choose the input state of the probe and, after a time $\tau$, to make a measurement of the probe state. 

Our setup is inspired by \cite{arXiv:1312.5724}, where Markovian master equations are studied with an added control field. Within this minimal scenario we show that it is possible to infer information about the unknown system, including properties of its state and spectrum, directly from the statistics of the final measurement of the probe. We find three universal regimes ordered in powers of $\lambda$, each of which witnesses different properties of the system. In particular, we show how to measure time-correlation functions for the system operators, which indicate the timescales over which the system behaves quantum mechanically. The simple probe that we describe here could be 
deployed to investigate a large variety of systems, and the more that can be assumed about a particular system, the more specific the information than can be obtained through our method, as we demonstrate with two examples -- one involving spin systems, the other concerning vibrational degrees of freedom.

%%%%%%%%%%%%%%%%%%%%%%%%
%%%%%%%%%%%%%%%%%%%%%%%%
\section{Perturbative expansion in the probe energy splitting}
%%%%%%%%%%%%%%%%%%%%%%%%
%%%%%%%%%%%%%%%%%%%%%%%%

\begin{figure}[t]
%\makebox[200pt]{
\includegraphics[width=200pt,viewport=0 0 200 92]%600pt 276pt]
{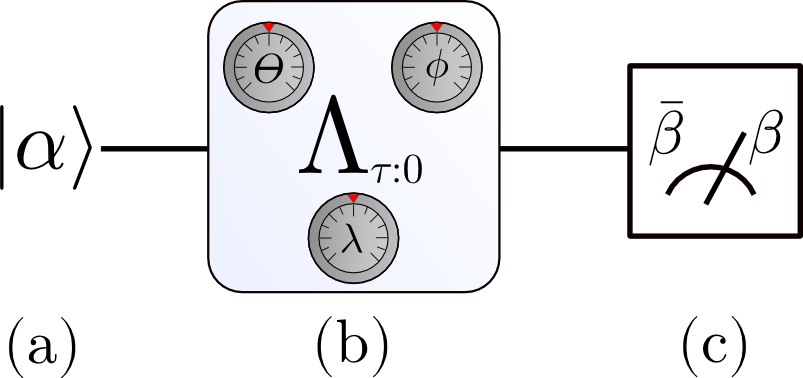}%}
\caption{(Color online) {\bf Probe evolution as a black box:}  (a) A two-level probe is prepared in state $\ket{\alpha}$; (b) the probe is acted on by CPTP map $\Lambda_{\tau:0}(\lambda,\theta,\phi)$, which has a set of controllable parameters corresponding to the magnitude and basis of an imposed splitting of the probe's energy levels $H_\s = \frac{1}{2}\lambda \, %\epsilon \,
\sigma_{(\theta,\phi)}$; (c) the probe is measured in the basis defined by state $\ket{\beta}$.} \label{fig:blackbox} \vspace{-10pt}
\end{figure}

Let us denote the prepared state of the probe as $\ket{\alpha}$ (our results can be straightforwardly extended to mixed states by linearity) and define the basis in which we measure by the state $\ket{\beta}$. The experimentally accessible quantity (from which the dynamical map $\Lambda_{\tau:0}$ can be recovered) is the transition probability
\begin{gather}\label{stats}
p_{\beta:\alpha} = \avg{\beta \vert \Lambda_{\tau:0}(\ketbra{\alpha}{\alpha}) \vert \beta}.
\end{gather}
By determining this quantity for different preparations and measurements as the control parameters are varied -- as depicted in Fig.~\ref{fig:blackbox} -- information about the system can be recovered. Furthermore, as we will show, much of this information can be recovered using preparations, measurements, and the control field in a single basis, meaning that our scheme could be used even with probes that have a preferred measurement basis. 

The method presented here is general; however, its practicality will be determined by our ability to increase the energy splitting of the probe, as well as to make preparations and measurements.
In practice, this may require that the probe is prepared independently (away from the system) and brought into and out of contact with the system in a repeatable manner, such that the two are not interacting when preparations and measurements are made. Alternatively, the probe can be placed next to the system and prepared via control operations on the probe alone, which must act on a timescale faster than that set by the interaction with the system.
%If there is no way of separating the probe from the larger preparation and measurement apparatus, then the latter will also, in general%Otherwise, the experimental apparatus should be included in the description of $\e$ -- an explicit treatment of which would depend on the particular implementation.

Combining Eqs.~\eqref{Stine}~\&~\eqref{stats},
the probability of finding the probe in the state $\ket{\beta}$, after the open dynamics takes place, can be written as the expectation value of the projector $\ketbra{\beta} {\beta} \otimes \mathbbm{1}$ with respect to the time evolved density operator:
\begin{gather}
p_{\beta:\alpha} = \tr \left[ (\ketbra{\beta}{\beta}\otimes\mathbbm{1}) U (\ketbra{\alpha}{\alpha}\otimes\rho_\e) U^\dg\right], \label{eq:transitionprob}
\end{gather}
where $U = \exp\left\{-\rmi H_{\text{tot}} \tau\right\}$ is the time evolution operator for the joint probe-system. In general, due to the unknown nature of $V_\se$, it is not possible to directly calculate $U$. However, by rewriting it as $U =  \exp\{-\rmi \left( \frac{1}{2} %\epsilon 
\lambda  \,\sigma_{(\theta,\phi)} \otimes \mathbbm{1}  + V_\se \right) \tau\} = \exp\{-\rmi \left( \frac{1}{2}%\epsilon  
\, \sigma_{(\theta,\phi)} \otimes \mathbbm{1}  + V_\se /\lambda \right) \lambda \tau \}$, one can see that, in the limit of large $\lambda$, $V_\se$ acts as a perturbation to the measurement Hamiltonian, albeit evolved to an effective time $\lambda \, \tau$. We can then move into the interaction picture and and write down the transition probabilities as an expansion in $1/\lambda$. This relative scaling of the probe and interaction Hamiltonians is related to that proposed by Davies in order to derive a weak coupling master equation \cite{Davies1976}.

To simplify the subsequent calculation, we can always decompose $V_\se$ in the basis defined by the eigenstates of the control Hamiltonian, $\{\ket{\pi_0},\ket{\pi_1}\}$,
\begin{align}
V_\se =& \ketbra{\pi_0}{\pi_0}\otimes A_0^c + \ketbra{\pi_0}{\pi_1} \otimes B^c \nonumber \\ &+ \ketbra{\pi_1}{\pi_0} \otimes {B^c}^\dg + \ketbra{\pi_1}{\pi_1} \otimes A_1^c, 
\end{align}
with $A_0^c = {A_0^c}^\dg$ and $A_1^c ={A_1^c}^\dg$. This has the advantage of further specifying how different system operators couple to the probe. By varying the angles $\theta$ and $\phi$, and hence the basis in which the bias is applied, different system operators can be selected to couple to different probe states (see Appendix~\ref{sec:anglevary} for details). From this point on we drop the explicit basis dependence for the system operators (\emph{e.g.} $A_0^c=A_0$).

If we move into the interaction picture with respect to $H_0 = %\frac{\epsilon}{2} 
\frac{1}{2}\sigma_{(\theta,\phi)} \otimes \mathbbm{1} + ({1}/{\lambda}) \ketbra{\pi_0}{\pi_0} \otimes A_0 + ({1}/{\lambda}) \ketbra{\pi_1}{\pi_1}\otimes A_1$, then the corresponding time evolution operator has a perturbative expansion. The resulting Hamiltonian is $\tilde{H}_t =  \exp\{\rmi %\epsilon 
t\} \ketbra{\pi_0}{\pi_1}\otimes \tilde{B}_t +\exp\{-\rmi %\epsilon 
t\} \ketbra{\pi_1}{\pi_0}\otimes \tilde{B}^\dg_t$, where $\tilde{B}_t = \exp\{\rmi A_0 t /\lambda\} \, B \, \exp\{-\rmi A_1 t /\lambda\}$.

In the interaction picture, the time evolution operator can be expanded in a Dyson series as
$\tilde{U}_t = \mathbbm{1} + \left({1}/{\rmi \lambda}\right) \int_0^{\lambda t} \rmd s \, \tilde{H}_s -\left({1}/{\lambda^2}\right) \int_0^{\lambda t} \rmd s \, \int_0^s \rmd s' \, \tilde{H}_s \tilde{H}_{s'} + \mathcal{O} \left({1}/{\lambda^3}\right),$
which in turn leads to an evolution equation for the density operator $\tilde{\rho}_t = \tilde{U}_t \, \rho_0 \, \tilde{U}^\dg_t$, where

\begin{widetext}
\begin{align}
\tilde{\rho}_t =& \rho_0  + \frac{1}{\rmi \lambda}\int_0^{\lambda t} \rmd s [\tilde{H}_s,\rho_0] 
-\frac{1}{\lambda^2} \int_0^{\lambda t} \rmd s \, \int_0^s \rmd s' \, \left(\tilde{H}_s\tilde{H}_{s'} \rho_0 + \rho_0 \tilde{H}_{s'}\tilde{H}_s \right)  
+ \frac{1}{\lambda^2} \int_0^{\lambda t} \rmd s \, \int_0^{\lambda t} \rmd s' \, \tilde{H}_s \rho_0\tilde{H}_{s'} +\, \cdots %\mathcal{O}\left(\frac{1}{\lambda^3}\right). 
\label{eq:rhoexpansion}
\end{align}
The Schr\"odinger picture density operator is given by $\rho_t = \exp\{-\rmi \lambda H_0 t\} \, \tilde{\rho}_t \, \exp\{\rmi \lambda H_0 t\}$, and the transition probability after evolving for time $\tau$ can be expressed, up to second order in $1/\lambda$, in terms of interaction picture quantities as (see Appendices~\ref{sec:lambdaconvergence}~\&~\ref{sec:pderivation} for details  and for the the conditions of validity of the perturbative series)
\begin{align}
p_{\beta:\alpha}=& \tr \left[ \ketbra{\beta}{\beta}\otimes\mathbbm{1}\, \rme^{-\rmi H_0 \lambda \tau} \tilde{U}_{\tau} \ketbra{\alpha}{\alpha}\otimes\rho_{\e} \tilde{U}^\dg_{\tau} \rme^{\rmi H_0 \lambda \tau}\right] \nonumber \\
%%%%%%%%
\simeq & \frac{q_{\beta:\alpha}}{2}  + a_0^* a_1 b_0 b_1^*\zeta^{(0)}(\lambda) %\nonumber \\
%%%%%%%%
+ \frac{1}{ \lambda%\epsilon
} \left[ a_0 a^*_1
\avg{\sigma_{(\theta,\phi)}}_\beta \zeta^{(1)}(\lambda) + b_0^* b_1 \left( |a_0|^2 \xi_0^{(1)}(\lambda) -  |a_1|^2 \xi_1^{(1)}(\lambda)\right)\right]\nonumber \\
%%%%%%%%
&- \frac{1}{\lambda^2%\epsilon^2
} \left[ \frac{\avg{\sigma_{(\theta,\phi)}}_\beta}{2} \left( |a_0|^2 \xi_0^{(2)}(\lambda) -  |a_1|^2 \xi_1^{(2)}(\lambda) \right) %\nonumber \\
%%%%%%%%
+a_0^* a_1 b_0 b_1^* \zeta^{(2)}(\lambda) \right] + {\rm c.c.} + \mathcal{O}\left(\frac{1}{\lambda^3}\right), \label{eq:alphaexpansion}
\end{align}
where $q_{\beta:\alpha} = |b_0a_0^*|^2 + |b_1a_1^*|^2$, $a_k = \braket{\alpha}{\pi_k}, \, b_k = \braket{\beta}{\pi_k}$, and
\begin{align}
&\zeta^{(0)}(\lambda) = \rme^{\rmi %\epsilon 
\lambda \tau}\avg{\rme^{-\rmi A_0 \tau} \rme^{\rmi A_1 \tau}}_\e, \quad
\zeta^{(1)}(\lambda) = \avg{B}_\e -\rme^{\rmi %\epsilon 
\lambda \label{eq:zeta0} \tau}\avg{\rme^{\rmi A_0 \tau}B \, \rme^{-\rmi A_1 \tau}}_\e, \\
&\xi_0^{(1)}(\lambda) = \avg{\rme^{\rmi A_0 \tau} B^\dg \rme^{-\rmi A_0 \tau}}_\e -\rme^{\rmi %\epsilon 
\lambda \tau} \avg{\rme^{\rmi A_0 \tau}\rme^{-\rmi A_1 \tau} B^\dg}_\e, \;
\xi_1^{(1)}(\lambda) = \avg{\rme^{\rmi A_1 \tau}B^\dg \rme^{-\rmi A_1 \tau}}_\e -\rme^{\rmi %\epsilon 
\lambda \tau} \avg{B^\dg \rme^{\rmi A_0 \tau}\rme^{-\rmi A_1 \tau}}_\e, \label{eq:xi1}\\
&\xi_0^{(2)}(\lambda) = \avg{B B^\dg}_\e+\avg{\rme^{\rmi A_0 \tau}BB^\dg\rme^{-\rmi A_0 \tau}}_\e -\rme^{\rmi %\epsilon 
\lambda  \tau}\avg{\rme^{\rmi A_0 \tau}B \, \rme^{-\rmi A_1 \tau}B^\dg}_\e,\label{eq:xi21}\\
&\xi_1^{(2)}(\lambda) = \avg{B^\dg B}_\e+\avg{\rme^{\rmi A_1 \tau}B^\dg B \, \rme^{-\rmi A_1 \tau}}_\e -\rme^{\rmi %\epsilon
\lambda \tau}\avg{B^\dg\rme^{\rmi A_0 \tau} B \rme^{-\rmi A_1 \tau}}_\e. \label{eq:xi22}
\end{align}
\end{widetext}
Above $\avg{\,\cdot\,}_\e = \tr[\;\cdot \,\rho_\e]$ and the functions $\{ \zeta^{(i)}(\lambda), \xi^{(i)}_k(\lambda) \}$ all depend on $\tau$, $V_\se$ and $\rho_\e$; their magnitudes remain constant with $\lambda$. $\zeta^{(2)}(\lambda)$ is not written out explicitly here, as it never appears at leading order in $\lambda$. It should be noted that Eq.~\eqref{eq:alphaexpansion} is linear in the choice of preparation and measurement state, and can thus be straightforwardly generalised to mixed initial states $\rho_\alpha$ and general measurement, \emph{positive operator valued measure}, elements $\Pi_\beta$ by making the replacements $a_k = \tr[\rho_\alpha \ketbra{\pi_k}{\pi_k}]$ and $b_k = \tr[\Pi_\beta \ketbra{\pi_k}{\pi_k}]$ respectively. These more general choices would introduce uncertainties to the procedure outlined below.

In Appendix~\ref{sec:lambdaconvergence}, we discuss in detail the conditions required of $\lambda$ for Eqs.~(\ref{eq:alphaexpansion}--\ref{eq:xi22}) to accurately describe the transition probabilities. In short, we require that $\lambda%\epsilon
\gg |\bra{j_1}B\ket{k_0}|$, where $\ket{j_1}$ and $\ket{k_0}$ are eigenstates of $A_0$ and $A_1$ respectively that have support in -- or are easily reached from -- the initial state $\rho_\e$. It is also necessary that the probe is not resonant with any energy splittings in the system that can be reached within three applications of the perturbation Hamiltonian $\tilde{H}_s$ (or at least that such resonances satisfy $|\bra{j_1}B\ket{k_0}|\lesssim 1/(\lambda^2 \tau)$). 

There is an important further restriction that depends on the leading order $\lambda$ dependence in Eq.~\eqref{eq:alphaexpansion} and which generally relates the value of $\lambda$ to the evolution time $\tau$. For zeroth and second order dependence, the constraints are
\begin{align}
\lambda %\epsilon 
\gg \sum_{j k}\kappa_{jk}^0|\bra{j_0}B^\dg B\ketbra{k_0}{k_0}\rho_\e\ket{j_0}| \label{eq:constraint0}
\end{align}
and
\begin{align}
\lambda %\epsilon 
\gg \sum_{jk}\kappa_{jk}^0\frac{|\bra{j_0}B^\dg B\ketbra{k_0}{k_0} B^\dg\rho_\e B\ket{j_0}|}{|\tr\{B^\dg \rho_\e B\}|}\label{eq:constraint2}
\end{align}
respectively, where 
\begin{align}
\kappa_{jk}^l= \begin{cases} 1/|E^l_j-E^l_{k}| \; \;& E^l_j \neq E^l_{k}\\
\;\tau \;\;& E^l_j = E^l_{k}
\end{cases}.
\end{align}
Here $E_j^0$ and $E_j^1$ are the eigenenergies of $A_0$ and $A_1$ respectively., i.e., $A_0 = \sum_j E_j^0 \ketbra{j_0}{j_0}$ and $A_1 = \sum_j E_j^1 \ketbra{j_1}{j_1}$.
The same expressions should also hold for $0\rightarrow 1$ and $B\leftrightarrow B^\dg$. When the transition probability is proportional to $1/\lambda$ to leading order, there is a similar constraint that is independent of $\lambda$:
\begin{align}
\tr\{B \rho_\e\} \gg \sum_{jk}\kappa_{jk}^0|\bra{j_0}B^\dg B\ketbra{k_0}{k_0}\rho_\e\ket{j_0}|,\label{eq:constraint1}
\end{align}
and the same with $0\rightarrow 1$ and $B\leftrightarrow B^\dg$.

With the exception of the latter, first order case, and practical limitations notwithstanding, sufficiently increasing the value of $\lambda$ will always lead to probabilities consistent with Eq.~\eqref{eq:alphaexpansion}. Even though the signature oscillations of $p_{\beta:\alpha}$ with $\lambda$ arise for energy splittings inconsistent with the constraints in Eqs.~\eqref{eq:constraint0}~\&~\eqref{eq:constraint2}, their validity could in principle be checked by further increasing $\lambda$ and again measuring the transition probabilities. Once the amplitudes of oscillation have converged, one can be sure that Eq.~\eqref{eq:alphaexpansion} correctly describes the dynamics.

\begin{table}[b]
\begin{tabular}{c||cc}
\hline 
$X$ & $ \ket{\beta} \notin \{\ket{\pi_0}, \ket{\pi_1}\} $   & $ \ket{\beta} \in \{\ket{\pi_0}, \ket{\pi_1}\} $ \\
\hline \hline 
$ \ket{\alpha} \notin \{\ket{\pi_0}, \ket{\pi_1}\} $   & $a_0^*a_1 b_0 b_1^* \, \zeta^{(0)} (\lambda)$ & $\pm a_0a_1^* \, \zeta^{(1)} (\lambda) / \lambda%\epsilon
$ \\
 $ \ket{\alpha} = \ket{\pi_k} $  & $(-1)^{k}b_0^* b_1 \, \xi_k^{(1)}(\lambda)/\lambda\epsilon$ & \;$(-1)^{k+1}\xi_k^{(2)}(\lambda)/\lambda^2%\epsilon^2
 $    
\end{tabular}
\caption{\textbf{Leading order $\lambda$ dependence of $p_{\beta:\alpha}$.}  The functional dependence is always oscillatory, with amplitude and phase that depend on the quantity $X$ given in Eq.~\eqref{eq:leadingorderp}, and an envelope function which depends on the order. The choice of preparation ($\ket{\alpha}$) and measurement ($\ket{\beta}$) determines $X$, and hence the dependence of $p_{\beta:\alpha}$ on $V_\se$ and $\rho_\e$. \label{tab:summary}}
\end{table}

%%%%%%%%%%%%%%%%%%%%%%%%
%%%%%%%%%%%%%%%%%%%%%%%%
\section{Inferring properties of the system}
%%%%%%%%%%%%%%%%%%%%%%%%
%%%%%%%%%%%%%%%%%%%%%%%%

We can see from Eq.~\eqref{eq:alphaexpansion} that the dominant term at large $\lambda$ depends on the choice of $\ket{\alpha}$ and $\ket{\beta}$: in general, zeroth order terms dominate, but when either the preparation or measurement is made in the $\{\ket{\pi_0}, \ket{\pi_1}\}$ basis, the leading order contribution is proportional to $1/\lambda$. When both $\ket{\alpha}$ and $\ket{\beta}$ are eigenstates of $H_\s$, then only terms that are second order in $1/\lambda$ survive. In each case, however, the probability for large $\lambda$ -- subject to the conditions discussed in the previous section -- can always be written in the form $p_{\beta:\alpha} \simeq q_{\beta:\alpha} + 2 \Re\{ X \}$, where the dependence of $X$ on the choice of preparation and measurement is summarised in Table~\ref{tab:summary}. 
The real part of $X$ is an oscillatory function of $\lambda$:
\begin{gather}\label{eq:leadingorderp}
2 \, \Re\{X\} \simeq \eta + D \cos(%\epsilon 
\lambda \tau + \varphi),
\end{gather}
with frequency $%\epsilon 
\tau$, an amplitude $D$, phase $\varphi$, and a shift $\eta$. All of $D$, $\varphi$ and $\eta$ depend on the system state and Hamiltonian.

This method for inferring properties of the system bears analogies with scattering theory. In both cases, a known wave function is used to probe the structure of an unknown object, resulting in phase shifts and amplitude/probability changes. By contrast, in the case investigated here, the probe is always coupled to its environment, so one has to find a way to switch off the interaction or, alternatively, to uncouple the two systems. Also, rather than looking at typical scattering quantities, such as the cross section, we focus on other structural properties of the system and the interaction, such as its survival probability in the initially prepared state. 

\begin{figure}[ht]
\begin{centering}
%\makebox[240pt]{
\includegraphics[width=240pt]%,bb=0pt 0pt 240pt 170pt]%720pt 489pt]
{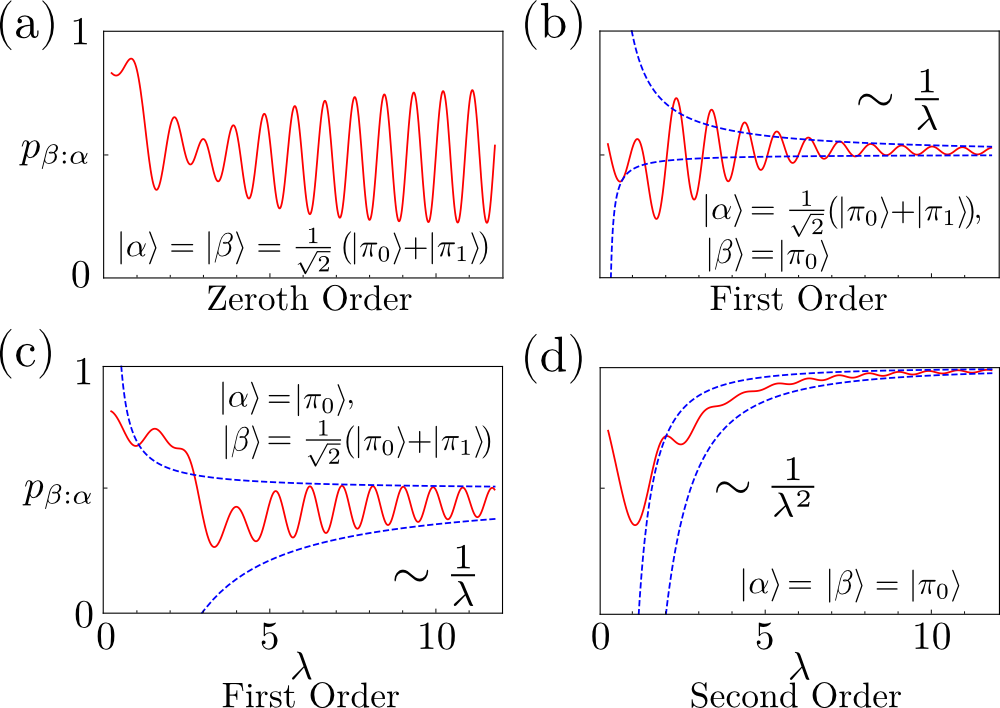}%}
\caption{(Color online) {\bf Transition probability.} Typical behaviour of $p_{\beta:\alpha}$ as $\lambda$ is varied from small to large (for reference the matrix elements of $V_\se$ are sampled such that $|\bra{j_{0(1)}} B \ket{k_{0(1)}}|\lesssim \sqrt{2}$). The results shown are for a system comprising two qubits; $V_\se$ and $\rho_\e$ were sampled randomly, such that they were effectively unknown prior to the simulations. Different panels show effects at different orders for large $\lambda$, corresponding to different choices of preparation and measurement. Panels (b) and (c) both correspond to first order effects, but with different choices of preparation. Solid red represent the values of $p_{\beta:\alpha}$, and bounding envelopes are shown with dashed blue lines.} \label{fig:dynamics}\vspace{-10pt}
\end{centering}
\end{figure}

In Fig.~\ref{fig:dynamics}, we plot $p_{\beta:\alpha}$ as a function of $\lambda$ for several different choices of preparation and measurement (with a randomly chosen $V_\se$ and $\rho_\e$ for a four-dimensional system). The probability can be seen to converge on a single oscillatory component with an appropriate envelope function. By characterising these oscillations, the values of the expectation values appearing in each of the terms of Eqs.~(\ref{eq:zeta0}-\ref{eq:xi22}) can be determined. If we further allow to vary the length of time for which the probe interacts with the system, then we can determine quantities such as $\avg{\rme^{-\rmi A_0 \tau} \rme^{\rmi A_1 \tau}}_\e$ and $\avg{\rme^{\rmi A_k \tau} B^\dg \rme^{-\rmi A_k \tau}}_\e$, as a function of $\tau$. We now discuss which elements of the system's energy spectrum and state can be recovered from these measurements, for more detail, see Appendix~\ref{sec:spectrum}.

When neither the preparation nor the measurement is in the basis of $H_\s$, the terms of Eq.~\eqref{eq:leadingorderp} become $\eta=0$ and $D \rme^{\rmi \varphi} \propto \avg{\rme^{-\rmi A_0 \tau} \rme^{\rmi A_1 \tau}}_\e$, with a coefficient that depends on $\ket{\alpha}$ and $\ket{\beta}$. Thus, the zeroth order term measures how differently the two eigenstates of the probe couple to the system. Oscillations that depend on system operators $A_0$ and $A_1$ can be attributed to `Lamb-shift' terms in the probe's Hamiltonian: an effective level splitting arising from its interaction with the system. The zeroth order becomes trivial, i.e., the oscillations no longer depend on the system when $A_0 = A_1$, i.e., when there is no interaction term proportional to $H_\s$ (in this case $A_0 = H_\e$, the Hamiltonian of the system). 

When either $\ket{\alpha}$ or $\ket{\beta}$ (but not both) are eigenstates of $H_\s$ and Eq.~\eqref{eq:constraint1} is satisfied, all zeroth order $\lambda$ dependences vanish and terms proportional to $1/\lambda$ dominate the large-$\lambda$ behaviour. From Table~\ref{tab:summary}, it can be seen that there are three distinct cases in which first order effects dominate, each measures a different property of the system. In particular, the Fourier components of the quantities $\avg{\rme^{\rmi A_k \tau} B^\dg \rme^{-\rmi A_k \tau}}_\e$ have frequencies equal to energy differences between eigenstates of $A_k$ which are initially in coherent superposition. Furthermore, when the central value $\eta$ of $\lambda$-oscillations dictated by $\xi^{(1)}_0$ or $\xi^{(1)}_1$ is independent of $\tau$ for some choice of $\theta$ and $\phi$, then the system state $\rho_\e$ is diagonal in the basis of the corresponding operator, $A_0$ or $A_1$ respectively. Remarkably, when $A_0=A_1$, oscillations at this order can therefore be used to test whether a system is in equilibrium.

Finally, when both preparations and measurements are made in the basis of $H_\s$, only second-order and higher terms survive in Eq.~\eqref{eq:alphaexpansion}. In this case, the system quantities that can be inferred are all two-time correlation functions for the operators $\tilde{B}_t$. These two-time correlation functions (also known as bath or reservoir correlation functions) dictate how the system influences the probe's dynamics in the absence of the control bias, quantifying memory effects and dynamical coherence \cite{NitzanBook, petruccione}. The second order effect can be related to the quantum Zeno effect \cite{zeno, FacchiPascazioZeno} as shown in \cite{arXiv:1312.5724}. The Fourier components of the correlation functions correspond to energy differences between all eigenstates of $A_0$ and $A_1$ that have support in $\rho_\e$ and those to which they couple through $B$. In other words, we get insight into the spectrum of $V_\se$ -- for a closed system this will be a series of discrete levels, whereas a system which is in turn coupled to a large reservoir will have a continuous distribution.  Lastly, the decay of $\lambda$-dependent oscillations at the second order (and hence the correlation function) above a certain evolution time $\tau$ signify a point beyond which system memory effects do not persist. Thus we can quantitatively measure the forgetfulness of a system, which is a generic property for large numbers of degrees of freedom.

Qualitatively, this behaviour is generic for quantum systems -- even when the system is further coupled to some environment. In this case the $B$ operators will couple subspaces of the larger closed system composed of $\e$ and its environment and, in many cases, there will be a continuum of energy spacings in the $A_0$ and $A_1$ operators. However, there are several exceptional cases that it is worth mentioning in more detail:

If the system $\e$ is classical and time-independent, $B \propto A_0 \propto A_1$ and the probe undergoes Hamiltonian dynamics; it must be considered as an isolated quantum system under the action of some potential. Since the potential is ``classical'', it will exhibit no recoil effects, and as a consequence, the quantum dynamics of $\s$ will not be able to unveil any interesting quantum-like feature of (classical) $\e$.

The same considerations would apply for a ``slowly'' changing classical system $\e$. In such a case the Hamiltonian of the probe $\s$ becomes time-dependent, and our results would still hold with all the $A$'s and $B$'s replaced by scalar functions of time.

On the other hand, a classical \emph{stochastic} system $\e$ would induce (unital) CPTP dynamics for the probe, by virtue of the Stinespring dilation theorem: the dynamics would be governed by a Gorini-Kossakowski-Sudarshan-Lindblad (self-dual) master equation, with potentially time-dependent rates. 
Our method would then yield the two-time correlation functions for $B$, which is now a stochastic variable. For more general semi-group dynamics, there is always a consistent time-independent Hamiltonian for a larger system that would produce the same behaviour.

%%%%%%%%%%%%%%%%%%%%%%%%
%%%%%%%%%%%%%%%%%%%%%%%%
\section{Example systems}
%%%%%%%%%%%%%%%%%%%%%%%%
%%%%%%%%%%%%%%%%%%%%%%%%

Of course, it is rare that a system is completely uncharacterised. More often than not, there is an assumed Hamiltonian for the system, which may include unknown parameters. It is also common to assume that the system is in thermal equilibrium with its environment. In such cases, assumptions about the system can be used to extract more specific information from the correlation functions discussed above. We now briefly present two generic and commonly used models for complex systems as examples, namely those comprising collections of spins and collections of harmonic oscillators (representing the vibrational modes of a molecule). The former tend to be used when the relevant degrees of freedom are highly localised \cite{spinbath}, whereas the latter are a good approximation to systems with delocalised degrees of freedom close to equilibrium \cite{FeynmanVernon1963}. 

%%%%%%%%%%%%%%%%%%%%%%%%
%%%%%%%%%%%%%%%%%%%%%%%%
\subsection*{Spin systems}
%%%%%%%%%%%%%%%%%%%%%%%%
%%%%%%%%%%%%%%%%%%%%%%%%

As our first example, we will consider a system comprising a small collection of spins -- for example molecular nuclear spins or isotopic impurities in otherwise spinless solids (for example silicon  \cite{TyryshkinMorton2006}). The energetic structure of such systems is routinely studied using nuclear magnetic resonance (NMR) spectroscopy, and characteristic spectra are used to identify different molecular species \cite{nmrbook}. However, obtaining NMR spectra usually requires a large ensemble of sample systems and strong magnetic field gradients. We will now demonstrate that our method could in principle be used to obtain spectra for single molecules using, for example, a tuneable magnetic dipole as a probe, without precise knowledge of the relative location of the sample with respect to the probe. It should be noted that our approach is not unique in this respect; other proposals have been put forward for single molecule NMR spectroscopy \cite{PerunicicHall2014}. A typical example of such a tuneable magnetic dipole would be a superconducting qubit, such as a flux qubit, although such probes are often fairly strongly coupled to wider environments. In practice, this coupling leads to rather short coherence times for flux qubits --  state-of-the-art $T_1$ and $T_2$ times are $\sim 40 \mu {\rm s}$ and $\sim 80 \mu {\rm s}$ respectively \cite{YanGustavsson2015} -- limiting the energy resolution of reconstructed spectra. What follows should be viewed, in the context of current technology, as a proof of principle for our scheme, rather than a concrete experimental simulation.

\begin{figure}[t]
%\makebox[200pt]{
\includegraphics[width=215pt]%,bb=0pt 0pt 200pt 92pt]%600pt 276pt]
{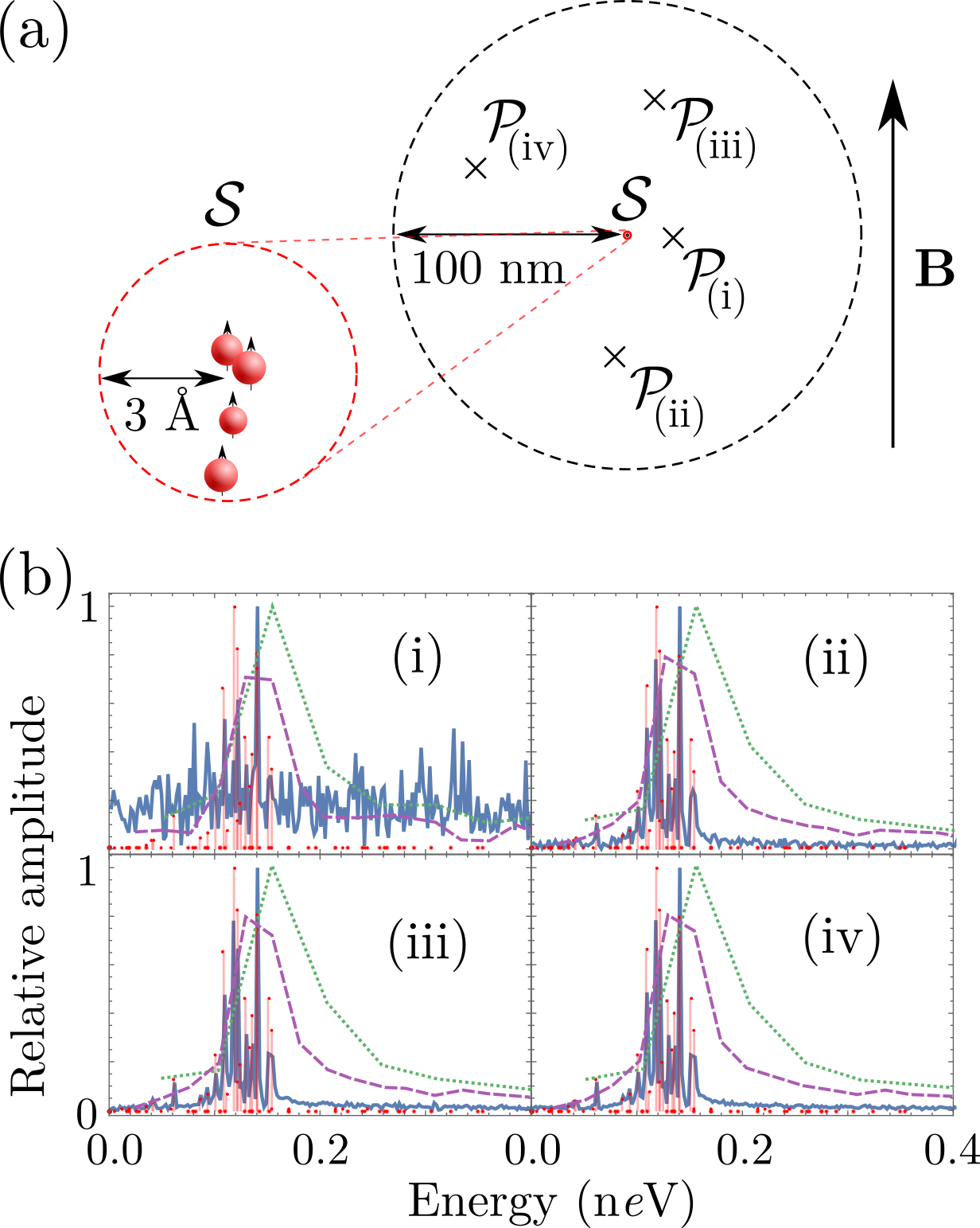}%}
\caption{(Color online) {\bf Simulated spin spectra:}  (a) A superconducting flux qubit probe $\s$ (with magnetic moment $0.3 \, {\rm meV \,T^{-1}}$) is placed at four different randomly chosen positions (i--iv) within a radius of $100\,{\rm nm}$ around a spin system $\e$, which  
consists of four spins (with magnetic moments $\sim \mu_N$) randomly arranged within a $0.01\,{\rm nm}$ sphere; a $|\mathbf{B}| = 1\,{\rm mT}$ magnetic field is applied perpendicular to the magnetic moment of $\s$. Probe positions are shown in the plane defined by $\mathbf{B}$ and the vector joining $\s$ and $\e$.  (b) Reconstructed spectra for the different probe positions. Red peaks indicate the spectrum that would be obtained with perfect measurement of the transition probabilities $p_{1:0}$ for all values of $\lambda$ and $\tau$. The other curves are simulated reconstructions, where the experiment was performed only a finite number ($\sim 10^6$) of times for each choice of parameters. $\tau$ was sampled evenly in $100 \rm ns$ steps for different total evolution times: $80 \,\rm \mu s$ (dotted green); $160 \,\rm \mu s$ (dashed purple); $2 \,\rm ms$ (solid blue). $100$ different values of $\lambda$ were tested in each case. \label{fig:nmr} }\vspace{-10pt}
\end{figure}

Spin systems in an external magnetic field $\mathbf{B}_0$ have a Hamiltonian comprising a Zeeman splitting term and a magnetic dipole-dipole interaction: 
\begin{align}
H_{\rm spin} =& -\sum_k \mu_k \mathbf{B}_0\cdot\mathbf{S}_k + \sum_{k k'} \frac{\mu_0 \mu_k \mu_{k'}}{4 \pi |\mathbf{r}_{kk'}|^3} \nonumber \\
&\qquad\!\!\times \left(\mathbf{S}_k \cdot \mathbf{S}_{k'}-\frac{(\mathbf{S}_k \cdot\mathbf{r}_{kk'})(\mathbf{S}_{k'} \cdot\mathbf{r}_{kk'})}{|\mathbf{r}_{kk'}|^2}\right),\label{eq:spinham}
\end{align}
where $\mathbf{r}_{kk'}=\mathbf{r}_k-\mathbf{r}_{k'}$ and  $\mathbf{S}_k$, $\mathbf{r}_k$ and $\mu_k$ are the spin operator, position and magnetic moment of the $k$th spin ($\mu_0$ is the permeability of free space). It is the differences between the eigenvalues of Eq.~\eqref{eq:spinham} -- usually broadened by the surrounding environment -- that determine the positions of peaks in NMR spectra. For nuclear spins in small molecules, the magnetic moment is on the order of the nuclear magneton $\mu_N$ and the nuclei are often only a few angstroms apart \cite{nmrbook}. In the absence of an external field, this leads to typical energy splittings of around $10\,{\rm peV}$.

The qubit probe itself is a magnetic dipole with magnitude $\mu_\s$, which is a function of the persistent current running through it; the orientation of this dipole moment depends on the quantum state of the qubit. If we use the two states of definite magnetic moment to define the eigenstates of the probe's $\sigma_x$ operator, then we can write its interaction with the spins as
\begin{align}
H_{\s \rm spin} =&  \sum_{k } \frac{\mu_0 \mu_\s \mu_{k}}{4 \pi |\mathbf{r}_{\s k}|^3} \sigma_x \\\nonumber 
&\qquad\!\!\times \left( \hat{n}_\s \cdot \mathbf{S}_{k}-\frac{(\hat{n}_\s \cdot\mathbf{r}_{\s k})(\mathbf{S}_{k} \cdot\mathbf{r}_{\s k})}{|\mathbf{r}_{\s k}|^2}\right),\label{eq:spinham2}
\end{align}
where $\hat{n}_\s$ is a unit vector along the probe's magnetic moment when it is in the $+1$ $\sigma_x$ eigenstate and $r_{\s k}$ is the vector joining the probe and the $k$th spin. It is also possible to tune the energy splitting of the flux qubit in Ref.~\cite{Paauw2008} between the eigenstates of $\sigma_z$ such that the persistent current (and hence magnetic moment) does not change appreciably, that is it satisfies the conditions required of a probe for our scheme.
 
Since only transitions between orthogonal probe states are of interest in this scenario, preparation of the probe can be performed using the same kind of projective measurement used to determine which of the two eigenstates of $\sigma_z$ the probe is in. For flux qubits, such measurements can be performed on timescales of a few $\rm ns$ -- much shorter than the interaction timescale between probe and system -- by biasing the energy of the definite current states (eigenstates of $\sigma_x$) and coupling it to, e.g., a Josephson bifurcation amplifier \cite{Knee2016,Nakano2009}.

In Fig.~\ref{fig:nmr}, we show a simulated reconstruction of the spectrum of a randomly positioned collection of spins in an external magnetic field, described by the Hamiltonian in Eq.~\eqref{eq:spinham}. The plots in the Figure are the Fourier transform of the two-time correlation functions in Eq.~\eqref{eq:xi21}, inferred using our scheme from measurements of the transition probability between orthogonal states of the probe; the different curves correspond to different total evolution times. Despite the fact that the probe is placed at different distances from the system, the same spectra are reconstructed, with greater or lesser accuracy. Only in those cases where the coupling is very strong, and the value of $\lambda$ necessarily very large, does the reconstruction fail; this is due to the difficulty of measuring small probabilities in a finite number of experiments. As can be seen from the Figure, the energy resolution of the reconstruction is limited by the evolution time of the joint probe-system; this is due to the comparatively long time scales associated with the spin system. In performing these simulations, we have assumed that other environments do not play a significant role in the dynamics on these time scales. While this is a poor assumption for flux qubits with current technology, where $1/f$ noise limits coherence times as discussed above, it is not unlikely that such experiments could be realistically performed in the not too distant future, given the rapid rate of improvement of isolation from their environment \cite{YanGustavsson2015}.

In general, other properties of the system can also be inferred by making preparations and measurements in different bases. We list the full set of obtainable quantities in Appendix~\ref{sec:spectrum}, where we show that when the system is highly disordered, the whole spectrum of the operators $A_0$ and $A_1$ can be identified. When there are no degenerate energy splittings, all diagonal elements of $\rho_\e$ in the basis of both $A_0$ and $A_1$ can also be determined.

%%%%%%%%%%%%%%%%%%%%%%%%
%%%%%%%%%%%%%%%%%%%%%%%%
\subsection*{Vibrational modes}\label{harmonic-oscillators}
%%%%%%%%%%%%%%%%%%%%%%%%
%%%%%%%%%%%%%%%%%%%%%%%%

Another possible application of the formalism presented here can be found in the context of molecular junctions, where electrons tunnelling between two nano-scale electronic binding sites (which could be small molecules or quantum dots) couple to the vibrational degrees of freedom of a molecule placed between them~\cite{nanoswitches}. The binding sites can be modelled as an electron donor and acceptor ($da$) pair -- a two level quantum system whose energy splitting can be tuned by varying the potential of leads to which they are connected; the intervening molecule then plays the role of the unknown system $\e$. Following Ref.~\cite{quantumnose}, we use a spin-boson model to describe the $da$ pair interacting with a molecule that has, along with its wider environment, a (continuous frequency) collection of vibrational modes:
\begin{align}
 H=&\frac{1}{2}\lambda%\epsilon_{da}
 \left(\ketbra{d}{d}-\ketbra{a}{a}\right)+V(\ketbra{d}{a}+\ketbra{a}{d})\nonumber\\
  % &+(\gamma_d\ketbra{d}{d}+\gamma_a\ketbra{a}{a})\otimes(b_0+b_0^\dagger)\nonumber\\
   &+ \sum_k(\gamma_{dk}\ketbra{d}{d}+\gamma_{ak}\ketbra{a}{a})\otimes(b_k+b_k^\dagger)\nonumber\\
   &%+\omega_{\e}b_0^\dagger b_0
   +\sum_k\omega_kb_k^\dagger b_k,
\end{align}
where the typical energy gap $\lambda% \epsilon_{da}
$ ($\sim100-200\,{\rm meV}$) is much larger than the tunneling energy $V$ ($\sim$1 meV) between donor and acceptor. To diagonalize the spin part of the Hamiltonian we use a polaron transformation \cite{McCutcheonNazir2011} generated by the operator 
$S = \sum_k (u_{dk} \ketbra{d}{d} - u_{ak} \ketbra{a}{a}) \otimes (b_k^\dagger -b_k)$, 
where $u_{Xk} = \gamma_{Xk} / \omega_{k}$. 

After performing the transformation, the energy gap is decreased by an amount $E_r=\sum_k\omega_k\left(u_{dk}^2-u_{ak}^2\right)$, known as the reorganisation energy, and the probe couples to the system through a $\sigma_x$ operator. 
If we additionally assume that the system and reservoir are both in a thermal state, we can then use Eq.~\eqref{eq:alphaexpansion} to write a simple formula for the transition probability from donor to acceptor (see Appendix~\ref{olf-details}):
\begin{align}
        p_{a: d} =& \frac{V^2}{\lambda^2% \epsilon^2
        }    \left[ 2 - \cos\left(\lambda%\epsilon
        \tau-E_r \tau\right)\,\exp\{ f(\tau)\} \right],\label{olf-probability}
\end{align}
where  
\begin{align}
    f(\tau)&=-\sum_j(u_{dj}-u_{aj})^2\coth\left(\frac{\omega_j}{2k_B T}\right) (1-\cos\omega_j\tau),\nonumber
\end{align}
in which $T$ is the temperature.

By measuring $p_{a:d}$ at different times and for different values of $\lambda$, we can both recover the reorganisation energy $E_r$ -- a measure of the overall coupling strength between $\s$ and $\e$ -- and reconstruct the function $f(\tau)$ in Eq.~\eqref{olf-probability}. Fourier transforming this function reveals information about the couplings $(\gamma_{kd}-\gamma_{ka})^2$, which for realistic systems can be written as a continuous distribution $J(\omega) = \sum_k (\gamma_{kd}-\gamma_{ka})^2 \delta(\omega-\omega_k)$, known as the spectral density. Explicitly,
\begin{align}
    \tilde{f}(\omega) = \frac{J(\omega)}{\omega^2} \coth\left(\frac{\omega}{2k_B T}\right),
\end{align}
for positive $\omega$. Not only can prominent vibrational modes of an unknown system $\e$ be identified by reconstructing $J(\omega)$, but for a system with known spectral density, $\tilde{f}(\omega)$ can be used to determine the temperature $T$. That is, in this case, our scheme can be used to do thermometry. 

In practice, the range of times for which one could reconstruct $f(\tau)$ is limited by the constraint in Eq.~\eqref{eq:constraint2}. For the parameters mentioned above, the maximum evolution time is $\sim 300\,{\rm ps}$ -- much longer than the vibrational coherence time for some typical spectral densities in large biomolecules~\cite{RitschelEisfeld2014}. If necessary, this time could be extended by increasing $\lambda$, since $\tau_{\rm max}\sim \lambda %\epsilon 
/ V^2$; however, this would reduce the transition probability $p_{a:d}$, which would then require more experiments to accurately measure. It should also be noted that other environments may contribute to the probe signal even in the absence of a sample system. For example, the leads which control the bias across the $da$-pair can be considered as a fermionic bath \cite{Brandes2005}. 

%%%%%%%%%%%%%%%%%%%%%%%%
%%%%%%%%%%%%%%%%%%%%%%%%
\section{Discussion}
%%%%%%%%%%%%%%%%%%%%%%%%
%%%%%%%%%%%%%%%%%%%%%%%%

We have presented a general scheme for using a two-level system to probe an unknown system to which it is coupled: the application of a precisely controlled external bias to the probe, in conjunction with measuring transition probabilities, allows for the reconstruction of many properties of the system, including information about its state and spectrum. Importantly, much of this information can be obtained with the probe being prepared and measured in a single preferred basis $\{\ket{\pi_0},\ket{\pi_1}\}$. 

Since our scheme rests on only a few assumptions -- notably we do not require the coupling strength to be in any particular regime, so long as the probe splitting can be made sufficiently large -- we expect it to find application in many physical scenarios. Our method can be seen as a kind of generalised spectroscopy, in that it gives us access to the spectrum of an unknown system. In particular, it could be used to determine the electronic and vibrational spectra of large molecular complexes, an important step in understanding many biological processes, such as photosynthesis \cite{ChinPrior2013}.

Unlike traditional spectroscopic techniques, however, the probe coupling in our approach is strictly off-resonant. In addition to satisfying the conditions in Eqs.~\eqref{eq:constraint0}~\&~\eqref{eq:constraint2}, which ensure the probe-system interaction acts perturbatively, the probe's splitting must be larger than energy gaps between system states which it directly couples.

Though the degree of control required to observe the oscillations in probability, and hence probe the system, is high, it is certainly within the realms of experimental possibility for some systems. In particular, ultrafast spectroscopy routinely investigates evolution on timescales of order $\sim\rm ns$ with a resolution of $\sim\rm fs$ or even $\sim\rm as$ \cite{LepineIvanov2014}. A comparatively long evolution time of $1 \rm ns$ requires control over the applied bias on the order of $\sim \rm \mu eV$ -- though, depending on the probe-system coupling, the magnitude of the bias may need to be much larger (see Appendix~\ref{sec:lambdaconvergence}). For a magnetic field splitting a hyperfine level of an alkali atom, this corresponds to a field strength of order $\sim 100 \rm m T$. Longer evolution times require an even greater degree of control, and larger values of $\lambda$ -- the maximum value of which may be fundamentally or technically limited in some systems -- to observe oscillations, but for many interesting physical processes, coherent quantum dynamics occurs on sufficiently short timescales. 

An interesting potential extension of our results would be to use the qubit probe to control a more complex system $\e$. For example, by regularly removing excitations from the probe (using some external time-dependent control), the latter could act as an energy sink for the system. While the interaction with the probe would allow it to escape `dark' subspaces (from which it cannot be directly cooled), the small effective coupling in the large-$\lambda$ regime would prevent it from being directly affected by the external control on the probe. The net effect would be to cool the degrees of freedom belonging to $\e$.

Finally, there are analogies but also important differences between the ideas outlined in this article and quantum state and process tomography. In quantum state tomography \cite{Bertrand1987}, one endeavours to make statements about the state of a given quantum system by performing (generalized)  measurements on the latter. These measurements are done by a classical apparatus or a field. On the other hand, in quantum process tomography \cite{nielsen, Poyatos1997, Modi2012SciRep} known quantum states are used to probe a quantum process and describe the quantum dynamics. In the approach outlined in this article, we are not varying the input and the output states, as one would do in a tomographic setting. The initial and final states are fixed, as is the probe Hamiltonian. We use the quantum system $\s$ to probe its environment $\e$ and then infer \textit{some} features of the latter from the induced dynamics of the former. Interestingly, different features are unearthed at different orders in the (inverse) energy splitting of $\s$. 

Clearly, one cannot expect a finite quantum probe to be able to unveil the full complexity of a much larger system with which it interacts. Nevertheless, memory effects, energy structure and coherent features do come to light, being associated with different couplings and different choices of initial states. It would be interesting to understand whether an exact solution of the problem (say, at all orders in the bias $\lambda$, if perturbation theory is consistent) would be able to probe different (deeper) characteristics of a complex system, including its most elusive quantum aspects.

\begin{acknowledgements}
The authors would like to thank Alex Monr\'as for useful discussions motivating this work. A.C. acknowledges financial support from the National Science Center, Sonata BIS Grant No. 2012/07/E/ST2/01402. S.P. was partially supported by the PRIN Grant No. 2010LLKJBX on ``Collective quantum phenomena: from strongly correlated systems to quantum simulators.".
\end{acknowledgements}

\appendix

\begin{widetext}
\section{Change of basis of control} 
\label{sec:anglevary}

The system operators for different choices of the control parameters can be related with reference to  their counterparts in the $z$-direction (which defines $\theta$ and $\phi$) as 
\begin{align}
A_0^c =& \cos^2\theta A^z_0 + \sin^2\theta A^z_1 + \cos\theta \sin\theta (B^z + {B^z}^\dg),  \\
A_1^c =& \sin^2\theta A^z_0 + \cos^2\theta A^z_1 - \cos\theta \sin\theta (B^z + {B^z}^\dg),  \\
B^c_{\hphantom{1}} =& \,\rme^{-\rmi \phi}\big[\sin^2\theta {B^z}^\dg - \cos^2\theta B^z + \cos\theta \sin\theta (A^z_0 - A^z_1)\big].
\end{align}

\section{Conditions for perturbation series to exist}
\label{sec:lambdaconvergence}
In order to only consider contributions at first and second order in $1/\lambda$, we require all higher order terms in Eq.~\eqref{eq:alphaexpansion}  to be strictly smaller in magnitude. Here we specify in detail the conditions on $\lambda$ for this to be the case.

Each term in Eq.~\eqref{eq:rhoexpansion} can be written in the form
\begin{align}
\frac{1}{(\rmi \lambda)^r} U^{(x)}_\tau \rho_0 {U^{(y)}_\tau}^\dg, \quad {\rm where} \quad
U^{(x)}_\tau = \int_0^{\lambda\tau} \rmd s_1\, \int_0^{s_1} \rmd s_2\, \cdots \int_0^{s_{x-1}} \rmd s_x\, \tilde{H}_{s_1}\tilde{H}_{s_2}\cdots \tilde{H}_{s_x} \label{eq:Uorderdef}
\end{align}
and $r=x+y$. Taking the trace over these terms, with the relevant projection operators for preparation and measurement states, leads to the following contributions to $p_{\beta:\alpha}$:
\begin{align}
&\frac{1}{(\rmi \lambda)^r}\tr\left[ \ketbra{\beta}{\beta}\otimes\mathbbm{1}\, \rme^{-\rmi H_0 \lambda \tau} U^{(x)}_{\tau} \ketbra{\alpha}{\alpha}\otimes\rho_{\e} {U^{(y)}_{\tau}}^\dg \rme^{\rmi H_0 \lambda \tau} \right] \nonumber \\
=& \frac{1}{(\rmi \lambda)^r}\sum_{\substack{nmpq%\in\{0,1\}
\\ jkll'}} a_n^*a_m b_p^*b_q \bra{j_n}\rho_{\e} \ket{k_m} \braket{l_p'}{l_q} e^{ (-1)^{p}\rmi \lambda %\epsilon 
\tau (1-\delta_{p,q})}\rme^{\rmi\tau(E_{l'}^p-E_l^q)}  \bra{\pi_q\,l_q}U_\tau^{(x)}\ketbra{\pi_n\,j_n}{\pi_m\,k_m} {U_\tau^{(y)}}^\dg\ket{\pi_p\,l'_p}, \label{eq:ordercontribution}
\end{align}
where we have expanded the system operators in their eigenbases as $A_0 = \sum_j E_j^0 \ketbra{j_0}{j_0}$ and $A_1 = \sum_j E_j^1 \ketbra{j_1}{j_1}$. Therefore, a valid perturbation expansion requires $\lambda\gg |\bra{\pi_q\,l_q}U_\tau^{(x)}\ketbra{\pi_n\,j_n}{\pi_m\,k_m} {U_\tau^{(y)}}^\dg\ket{\pi_p\,l'_p}|^{1/(x+y)}$. We can now use the definition of $U_\tau^{(x)}$ in Eq.~\eqref{eq:Uorderdef}, as well as the interaction Hamiltonian given in the main text to write these amplitudes explicitly. Firstly, note that 
\begin{align}
\bra{\pi_q\,l_q}U_\tau^{(x)}\ket{\pi_n\,j_n} \propto 
\begin{cases}
\delta_{n,q} & x\; {\rm even},\\
1-\delta_{n,q}\quad & x\; {\rm odd}.
\end{cases}\label{eq:oddevenU}
\end{align}
We will therefore treat these two cases separately. For the even case, we have
\begin{align}
\bra{\pi_0\,l_0}U_\tau^{(x)}\ket{\pi_0\,j_0} =& \sum_{k^{(1)}k^{(2)}\cdots k^{(x-1)}} \bra{l_0\vphantom{k^{(1)}_1}}B\ketbra{k^{(1)}_1}{k^{(1)}_1}B^\dg\ketbra{k^{(2)}_0}{k^{(2)}_0}\cdots\ketbra{k^{(x-1)}_1}{k^{(x-1)}_1}B^\dg\ket{j_0\vphantom{k^{(1)}_1}} \nonumber \\
&\qquad \times \int_0^{\lambda\tau} \rmd s_1\, \int_0^{s_1} \rmd s_2\, \cdots \int_0^{s_{x-1}} \rmd s_x\, \rme^{\rmi s_1 \left(1%\epsilon
+ (E^0_{l}-E^1_{k^{(1)}})/\lambda\right)} \rme^{-\rmi s_2 \left(1%\epsilon
+ (E^0_{k^{(2)}}-E^1_{k^{(1)}})/\lambda\right)} \nonumber \\
&\qquad\qquad\qquad\qquad\qquad\qquad\qquad\qquad\cdots \rme^{\rmi s_{x-1} \left(1%\epsilon
+ (E^0_{k^{(x-2)}}-E^1_{k^{(x-1)}})/\lambda\right)} \rme^{-\rmi s_x \left(1%\epsilon 
+ (E^0_{j}-E^1_{k^{(x-1)}})/\lambda\right)},
\end{align}
and 
\begin{align}
\bra{\pi_1\,l_1}U_\tau^{(x)}\ket{\pi_1\,j_1} =& \sum_{k^{(1)}k^{(2)}\cdots k^{(x-1)}} \bra{l_1\vphantom{k^{(1)}_1}}B\ketbra{k^{(1)}_0}{k^{(1)}_0}B^\dg\ketbra{k^{(2)}_1}{k^{(2)}_1}\cdots\ketbra{k^{(x-1)}_0}{k^{(x-1)}_0}B^\dg\ket{j_1\vphantom{k^{(1)}_1}} \nonumber \\
&\qquad \times \int_0^{\lambda\tau} \rmd s_1\, \int_0^{s_1} \rmd s_2\, \cdots \int_0^{s_{x-1}} \rmd s_x\, \rme^{-\rmi s_1 \left(1%\epsilon
+ (E^0_{k^{(1)}}-E^1_{l})/\lambda\right)} \rme^{\rmi s_2 \left(1%\epsilon
+ (E^0_{k^{(1)}}-E^1_{k^{(2)}})/\lambda\right)} \nonumber \\
&\qquad\qquad\qquad\qquad\qquad\qquad\qquad\qquad\cdots \rme^{-\rmi s_{x-1} \left(1%\epsilon
+ (E^0_{k^{(x-1)}}-E^1_{k^{(x-2)}})/\lambda\right)} \rme^{\rmi s_x \left(1%\epsilon
+ (E^0_{k^{(x-1)}}-E^1_{j})/\lambda\right)}.
\end{align}
For the odd case:
\begin{align}
\bra{\pi_1\,l_1}U_\tau^{(x)}\ket{\pi_0\,j_0} =& \sum_{k^{(1)}k^{(2)}\cdots k^{(x-1)}} \bra{l_1\vphantom{k^{(1)}_1}}B\ketbra{k^{(1)}_0}{k^{(1)}_0}B^\dg\ketbra{k^{(2)}_1}{k^{(2)}_1}\cdots\ketbra{k^{(x-1)}_1}{k^{(x-1)}_1}B\ket{j_0\vphantom{k^{(1)}_1}} \nonumber \\
&\qquad \times \int_0^{\lambda\tau} \rmd s_1\, \int_0^{s_1} \rmd s_2\, \cdots \int_0^{s_{x-1}} \rmd s_x\, \rme^{-\rmi s_1 \left(1%\epsilon
+ (E^0_{k^{(1)}}-E^1_{l})/\lambda\right)} \rme^{\rmi s_2 \left(1%\epsilon
+ (E^0_{k^{(1)}}-E^1_{k^{(2)}})/\lambda\right)} \nonumber \\
&\qquad\qquad\qquad\qquad\qquad\qquad\qquad\qquad\cdots \rme^{\rmi s_{x-1} \left(1%\epsilon
+ (E^0_{k^{(x-2)}}-E^1_{k^{(x-1)}})/\lambda\right)} \rme^{-\rmi s_x \left(1%\epsilon
+ (E^0_{j}-E^1_{k^{(x-1)}})/\lambda\right)}.
\end{align}
We can use these expressions to relate the magnitude of lambda required for a good perturbation series to the operators $A_k$ and $B$. In order for the contribution from the term proportional to $1/\lambda^r$ to be smaller in magnitude than that proportional to $1/\lambda^w$, if we ignore the time-dependent part it is sufficient that
\begin{align}
\lambda %\epsilon 
\gg |\bra{j_1}B\ket{k_0}|, \label{eq:lambdamods1}
\end{align}
for all $\ket{j_1}$ and $\ket{k_0}$ reachable from $\rho_\e$ within $w+1$ applications of the interaction Hamiltonian (i.e., $\bra{j_1(k_0)}\tilde{H}_0^{w+1}\rho_\e\tilde{H}_0^{w+1}\ket{j_1(k_0)}\neq0$), and
\begin{align}
\lambda %\epsilon 
\gsim |\bra{j_1}B\ket{k_0}|, \label{eq:lambdamods2}
\end{align}
for all $\ket{j_1}$ and $\ket{k_0}$ reachable from $\rho_\e$ within $r$ interactions.

When $\lambda%\epsilon
$ does not equal any of the energy splittings in the system (we will consider the effect of resonances below), iteratively performing the first few nested integrals leads to
\begin{align}
& \int_0^{\lambda \tau} \rmd s_1 \,\int_0^{s_1} \rmd s_2 \, \cdots \int_0^{s_{x-1}} \rmd s_x \, \rme^{-\rmi  s_1 \left(1%\epsilon
+ (E^0_{l} - E^1_{k^{(1)}} )/\lambda \right)} \rme^{\rmi  s_2 \left(1%\epsilon
+ (E^0_{k^{(2)}} - E^1_{k^{(1)}} )/\lambda \right)} \cdots \rme^{-\rmi  s_x\left(1%\epsilon
+ (E^0_{j} - E^1_{k^{(x-1)}} )/\lambda \right)} \nonumber \\
=& \rmi  \int_0^{\lambda \tau} \rmd s_1 \,\int_0^{s_1} \rmd s_2 \, \cdots \int_0^{s_{x-2}} \rmd s_{x-1} \, \rme^{-\rmi  s_1\left(1%\epsilon
+ (E^0_{l} - E^1_{k^{(1)}} )/\lambda \right)} \cdots \rme^{-\rmi  s_{x-2}\left(1%\epsilon
+ (E^0_{k^{(x-2)}} - E^1_{k^{(x-3)}} )/\lambda \right)} \nonumber \\
& \qquad \times \frac{\rme^{\rmi  s_{x-1}\left((E^0_{k^{(x-2)}} - E^1_{k^{(x-1)}})-(E^0_{j} - E^1_{k^{(x-1)}}) \right)/\lambda } - \rme^{\rmi  s_{x-1}\left(1%\epsilon
+ (E^0_{k^{(x-2)}} - E^1_{k^{(x-1)}} )/\lambda \right)}}{1%\epsilon
+ (E^0_{j} - E^1_{k^{(x-1)}} )/\lambda} \nonumber \\
=& -(\rmi)^2 \int_0^{\lambda \tau} \rmd s_1 \,\int_0^{s_1} \rmd s_2 \, \cdots \int_0^{s_{x-3}} \rmd s_{x-2} \, \rme^{-\rmi  s_1\left(1%\epsilon
+ (E^0_{l} - E^1_{k^{(1)}} )/\lambda \right)} \cdots \rme^{\rmi  s_{x-3}\left(1%\epsilon
+ (E^0_{k^{(x-4)}} - E^1_{k^{(x-3)}} )/\lambda \right)} \nonumber \\
& \qquad \times \Bigg(\lambda\frac{\rme^{-\rmi  s_{x-2}\left(1%\epsilon
+(E^0_{k^{(x-2)}} - E^1_{k^{(x-3)}})/\lambda-(E^0_{k^{(x-2)}} - E^1_{k^{(x-1)}})/\lambda+(E^0_{j} - E^1_{k^{(x-1)}})/\lambda \right) } - \rme^{-\rmi  s_{x-2}\left(1%\epsilon
+ (E^0_{k^{(x-2)}} - E^1_{k^{(x-3)}} )/\lambda \right)}}{\left( (E^0_{k^{(x-2)}} - E^1_{k^{(x-1)}})-(E^0_{j} - E^1_{k^{(x-1)}})\right)\left(1%\epsilon
+ (E^0_{j} - E^1_{k^{(x-1)}} )/\lambda\right)} \nonumber \\
& \qquad\qquad \qquad - \;\;\frac{\rme^{-\rmi  s_{m-2}\left((E^0_{k^{(x-2)}} - E^1_{k^{(x-3)}})-(E^0_{k^{(x-2)}} - E^1_{k^{(x-1)}}) \right)/\lambda } - \rme^{-\rmi  s_{m-2}\left(1%\epsilon
+ (E^0_{k^{(x-2)}} - E^1_{k^{(x-3)}} )/\lambda \right)}}{\left(1%\epsilon
+ (E^0_{j} - E^1_{k^{(x-1)}} )/\lambda\right)\left(1%\epsilon
+ (E^0_{k^{(x-2)}} - E^1_{k^{(x-1)}} )/\lambda\right)}  \Bigg).\label{eq:integrations}
\end{align}
As can be seen in the second to last line of Eq.~\eqref{eq:integrations}, factors of $\lambda$ appear in the numerator whenever the $1%\epsilon
$'s cancel in the exponent (one can see that factors of $1/\lambda$ in the denominator contribute to higher order terms in the expansion). For a given term, this will happen at most every other integration (\eg~when integrating over $s_{x-1}$, $s_{x-3}$, $s_{x-5}$ etc.), hence the leading order term of the whole integral will be explicitly proportional to $\lambda^{\frac{x}{2}}$ in the even case and $\lambda^{\frac{x-1}{2}}$ in the odd case. If we were to require the magnitude of such terms to still be strictly smaller than those at lower orders, we would need that the ratio of subsequent terms goes as $1/\lambda$. Ignoring the complex exponential factors resulting from the integration, this constraint can be written as
\begin{align}
\frac{|\tr\{(B B^\dg)^{x}\ketbra{l_0}{l_0} B B^\dg \ketbra{l'_0}{l'_0} (B B^\dg)^{r/2-x} B \rho_\e\}|/|E^0_l-E^0_{l'}|}{|\tr\{(B B^\dg)^{r/2} \rho_\e\}|} \lesssim &\frac{1}{\lambda}\; \forall \,E^0_l\neq E^0_{l'}\; \text{and} \; \forall \, x \leq r/2 \; (r \;\text{even}), \nonumber \\
\frac{|\tr\{(B B^\dg)^{x}\ketbra{l_0}{l_0} B B^\dg \ketbra{l'_0}{l'_0} (B B^\dg)^{(r+1)/2-x}  \rho_\e\}|/|E^0_l-E^0_{l'}|}{|\tr\{(B B^\dg)^{(r-1)/2}B \rho_\e\}|} \lesssim &\frac{1}{\lambda}\; \forall \,E^0_l\neq E^0_{l'}\; \text{and} \; \forall \, x \leq r/2 \; (r \;\text{odd}),\label{eq:strictconstraints}
\end{align}
and similar expressions with additional pairs of projectors sandwiched between $B$ operators and corresponding factors of $1/|E^0_k-E^0_{k'}|$. The same should also hold with $0\rightarrow 1$ and $B^\dg B \rightarrow B B^\dg$.

There will also be terms appearing in Eq.~\eqref{eq:integrations} where the pairs of energy levels in the exponent cancel, i.e., $(E^0_{l} - E^1_{k})-(E^0_{l'} - E^1_{k}) = E^0_{l}-E^0_{l'} = 0$. Even when there are no degeneracies, this can occur when $l=l'$. These terms introduce extra factors of $\lambda \tau$, leading to a restriction on the time for which the perturbation series remains valid. Interacting for a time $\tau$ places constraints on $\lambda$ identical to those in Eq.~\eqref{eq:strictconstraints}, but with $1/|E^0_l-E^0_{l'}|$ replaced by $\tau$ (and energies required to be equal, $E^0_l = E^0_{l'}$).

However, for the results in the main text to hold we do not need each subsequent term in Eq.~\eqref{eq:rhoexpansion} to be strictly smaller than the last, only that later contributions are much smaller than the zeroth, first or second order terms we present in Eqs.~(\ref{eq:zeta0}--\ref{eq:xi22}) (depending on the choice of preparation and measurement). For the zeroth order case, in addition to Eqs.~\eqref{eq:lambdamods1}~\&~\eqref{eq:lambdamods2}, we must also worry about contributions from second order terms of the sort discussed below Eq.~\eqref{eq:integrations}. Though they are nominally proportional to $1/\lambda$ after taking into account the integration, the coefficients could in principle be large; considering expressions similar to those in Eq.~\eqref{eq:strictconstraints}, where the second order terms are required to be much smaller than those at zeroth order, it is straightforward to show that sufficient constraints on $\lambda$ are
\begin{align}
\lambda %\epsilon 
\gg \sum_{l l'}\kappa_{ll'}^0|\bra{l_0}B^\dg B\ketbra{l'_0}{l'_0}\rho_\e\ket{l_0}|    \quad \text{where} \quad \kappa_{ll'}^j= \begin{cases} 1/|E^j_l-E^j_{l'}| \; \;& E^j_l \neq E^j_{l'}\\
\;\tau \;\;& E^j_l = E^j_{l'}
\end{cases},\label{eq:constraintzeroth}
\end{align}
and the same with $0\rightarrow 1$ and $B^\dg B \rightarrow B B^\dg$. 

For the first order terms in the main text to accurately describe the transition probabilities, we require that the second order terms are much smaller. This leads to $\lambda$--independent limits on $\rho_\e$ and the $A_k$ and $B$ operators:
\begin{align}
\tr\{B \rho_\e\} \gg \sum_{l l'}\kappa_{ll'}^0|\bra{l_0}B^\dg B\ketbra{l'_0}{l'_0}\rho_\e\ket{l_0}|,\label{eq:constraintfirst}
\end{align}
and the same with $0\rightarrow 1$ and $B^\dg B \rightarrow B B^\dg$, where $\kappa_{ll'}^j$ is defined as in Eq.~\eqref{eq:constraintzeroth}.

When both preparation and measurement are in the control basis, the corresponding constraints are again dependent on $\lambda$, since for the case $\braket{\beta}{\alpha}=0$ (trivially related to the case with $\braket{\beta}{\alpha}=1$), the next highest terms are at fourth order in the perturbation series ($r=4$ in Eq.~\eqref{eq:ordercontribution}). Strictly they are
\begin{align}
\lambda %\epsilon
\gg \sum_{l l'}\kappa_{ll'}^0\frac{|\bra{l_0}B^\dg B\ketbra{l'_0}{l'_0} B^\dg\rho_\e B\ket{l_0}|}{|\tr\{B^\dg \rho_\e B\}|},\label{eq:constraintsecond}
\end{align}
and the same with $0\rightarrow 1$ and $B \leftrightarrow B^\dg$.

In the case where there are energy differences $E^0_{j} - E^1_{k} = \lambda% \epsilon
$, then, in general, extra powers of $\lambda\tau$ appear from the nested integrals, and the perturbation expansion is no longer valid. However, if we restrict such energy differences to transitions which require at least $w$ interactions to reach from the initial system state, i.e., $|\bra{j_0}(\tilde{H}_{s})^z \rho_\e (\tilde{H}_{s})^z\ket{j_0}| \simeq |\bra{k_1}(\tilde{H}_{s})^z \rho_\e (\tilde{H}_{s})^z\ket{k_1}| \simeq 0$ for $z<w$, then terms in Eq.~\eqref{eq:ordercontribution} with explicit $1/\lambda^r$ dependence will include at most $r-2w$ additional powers of $\lambda$. Formally, we require $\lambda%\epsilon 
\neq |E^0_{j} - E^1_{k}|$ for all $\ket{k_1}$, $\ket{j_0}$, such that $\tr[\ketbra{j_0}{j_0} \tilde{H}_s \ketbra{k_1}{k_1} \tilde{H}_s^x \rho_\e \tilde{H}_s^{y}]\neq 0$, where $x+y\le w-1$.

For continuous frequency systems (often the case when a system of interest cannot be isolated from its wider environment), such resonances are unavoidable;  $\lambda %\epsilon 
= E^0_{j} - E^1_{k}$ always, for some $E^0_{j}$ and $E^1_{k}$. The couplings for such a system can be rewritten in terms of continuous functions $G(\omega) = \sum_{j,k} \bra{k_1} B \ket{j_0} \delta(\omega - |E^0_{j} - E^1_{k}|)$. Our perturbation expansion still holds when $G(\lambda% \epsilon
) \lsim 1/(\lambda^2\tau)$ -- effectively removing the offending terms to leading order.

With general $\ket{\alpha}$ and $\ket{\beta}$, all terms of the form given in Eq.~\eqref{eq:ordercontribution} contribute to $p_{\beta:\alpha}$, so the leading order contribution from the $1/\lambda^r$ term is actually $\lambda^{x/2}\lambda^{y/2}/\lambda^r=1/\lambda^{r/2}$ (x and y both even), $\lambda^{(x-1)/2}\lambda^{y/2}/\lambda^r=1/\lambda^{(r+1)/2}$ (one even, one odd) or $\lambda^{(x-1)/2}\lambda^{(y-1)/2}/\lambda^r=1/\lambda^{r/2+1}$ (x and y both odd), depending on the value of $r$ (and neglecting the resonance conditions described above). After those terms treated explicitly in the main text, terms with $r=1$ and $r=2$ have the largest contributions, but neither has contributions at leading (zeroth) order, thus they can be neglected. This remains the case as long as there are no energy gaps resonant with $\lambda%\epsilon
$ within $w=1$ interaction of $\rho_\e$.

In the case where $\ket{\alpha}$ or $\ket{\beta}$ is in the control basis, we must consider  whether terms with $r=2$ or greater contribute at first order in $1/\lambda$. Using Eq.~\eqref{eq:integrations}, we find that $r=2$ and $r=3$ terms are both proportional to $1/\lambda^2$ at leading order, as long as there are no $w=1$ resonances.

Finally, when $\ket{\alpha}$ and $\ket{\beta}$ are both in the control basis, we need to make sure $r\ge3$ terms do not contribute at second order. Though $r=3$ terms go to zero in this case, we might worry that for $r=4$, terms with $x=2$, $y=2$ in Eq.~\eqref{eq:ordercontribution} do have second order contributions. However, when $\braket{\alpha}{\beta}=0$ (which is trivially related to the case where they are identical), Eq.~\eqref{eq:oddevenU} implies that such terms disappear; therefore, without resonances, $r=4$ contributes at third order at most (with $x=1$ and $y=3$) and $r=6$ at fourth order. In order for there to be no contributing resonances, we require that $\lambda$ is greater than all energy differences within $w=3$ interactions.

To summarise: for the results in the main text to hold, we firstly require that $\lambda %\epsilon 
\gg \bra{j_0} B \ket{k_1}$, at least for those energy levels $\ket{j_0}$ and $\ket{k_1}$ which are within three interactions of the initial state $\rho_\e$ ($\bra{j_0}B B^\dg B \rho_\e B B^\dg B \ket{k_1}\neq0$). We also require that all other matrix elements $\bra{j_0} B \ket{k_1}$ are not significantly greater than $\lambda%\epsilon
$. Furthermore, we need all energy differences that involve aforementioned energy levels, reachable within three interactions, to be different from the probe splitting: $\lambda %\epsilon 
\neq |E_j^0-E_k^1|$. There is a final requirement given by Eqs.~(\ref{eq:constraintzeroth}--\ref{eq:constraintsecond}), depending on the leading order term of $p_{\beta:\alpha}$ (and hence choice of preparation and measurement) that relates $\lambda$ to the interaction time $\tau$ and the properties of the system for zeroth or second order leading $1/\lambda$ dependence. For first order dependence, there is a fundamental constraint on the properties of the system.

\section{$\lambda$-dependence of $p_{\beta:\alpha}$}
\label{sec:pderivation}

\emph{Order one---} By expanding the system operators in their eigenbases as $A_0 = \sum_n E_n^0 \ketbra{n_0}{n_0}$ and $A_1 = \sum_n E_n^1 \ketbra{n_1}{n_1}$, we can perform the integrals in Eq.~\eqref{eq:rhoexpansion}. For the first order term, we have
\begin{align}
& \frac{1}{\rmi\lambda} \int_0^{\lambda \tau} \rmd s \,\Big\{  \tr \big[a_0a_1^*\avg{\sigma_{(\theta,\phi)}}_\beta \rme^{i %\epsilon 
s}\tilde{B}_s \rho_\e  - b^*_0 b_1\rme^{-i %\epsilon 
(s-\lambda\tau)}
\tr \big[\big(|a_1|^2\rho_{\e} \tilde{B}^\dg_s  - |a_0|^2 \tilde{B}^\dg_s \rho_{\e} \big)\rme^{\rmi A_0 \tau}\rme^{-\rmi A_1 \tau}\big]\Big\}+{\rm c.c.} \nonumber \\
=& -\frac{1}{\lambda%\epsilon 
} \sum_{nm}\frac{1-\rme^{-i(\lambda %\epsilon
+(E_n^0-E_m^1)) \tau}}{1+(E_n^0-E_m^1)/\lambda%\epsilon
} \bigg(b^*_0b_1\sum_{l}|a_1|^2 \bra{m_1}B^\dg \ket{n_0}\braket{n_0}{l_1}\bra{l_1}\rho_\e \ket{m_1}\rme^{i(\lambda %\epsilon
+(E_n^0-E_l^1)) {\tau}} \nonumber \\
&\quad  -b^*_0 b_1\sum_{l}|a_0|^2\bra{n_0}\rho_\e \ket{l_0}\braket{l_0}{m_1}\bra{m_1}B^\dg \ket{n_0}\rme^{i(\lambda %\epsilon
+(E_p^0-E_m^1)) \tau} - a_0a^*_1\avg{\sigma_{(\theta,\phi)}}_\beta \bra{n_0}B \ketbra{m_1}{m_1}\rho_\e \ket{n_0}
\bigg) +{\rm c.c.}\nonumber \\
=&\frac{1}{\lambda%\epsilon
} \Big\{a_0 a_1^* \avg{\sigma_{(\theta,\phi)}}_{\beta} \big(\avg{B}_\e - \avg{\rme^{\rmi A_0 \tau} B \rme^{-\rmi A_1 \tau}}_\e\,\rme^{\rmi \lambda %\epsilon 
\tau}\big)  + b_0^*b_1|a_0|^2\big( \avg{\rme^{\rmi A_0 \tau} B^\dg \rme^{-\rmi A_0 \tau}}_\e- \avg{\rme^{\rmi A_0 \tau}\rme^{-\rmi A_1 \tau} B^\dg}_\e \rme^{\rmi \lambda %\epsilon
\tau}\big) \nonumber \\
&\qquad\qquad\qquad\;\; - b_0^*b_1 |a_1|^2 \big(\avg{\rme^{\rmi A_1 \tau} B^\dg \rme^{-\rmi A_1 \tau}}_\e-\avg{B^\dg \rme^{\rmi A_0 \tau}\rme^{-\rmi A_1 \tau}}_\e \rme^{\rmi \lambda %\epsilon
\tau}\big)\Big\}+{\rm c.c.} +\mathcal{O}\left(\frac{1}{\lambda^2}\right), \label{eq:firstorder}
\end{align}
where in the final line we see $\zeta^{(1)}(\lambda)$, $\xi^{(1)}_0(\lambda)$ and $\xi^{(1)}_1(\lambda)$ as defined in Eqs.~\eqref{eq:zeta0}~\&~\eqref{eq:xi1} appearing. The second order (and higher) terms that result from expanding the denominator in the second line are absorbed into the definition of $\zeta^{(2)}$ in Eq.~\eqref{eq:alphaexpansion}.

\emph{Order two---} Similarly, we can perform the integrals in the second order terms in Eq.~\eqref{eq:rhoexpansion}. We ignore those terms which never appear at leading order in $1/\lambda$, which are included in the definition of $\zeta^{(2)}(\lambda)$. For the remaining terms we find
\begin{align}
-\frac{1}{ 2\lambda^2} &\int_0^{\lambda \tau} \rmd s \, \int_0^{\lambda \tau} \rmd s' \, \tr\Big[ \avg{\sigma_{(\theta,\phi)}}_\beta \Big( |a_0|^2 \rme^{\rmi %\epsilon 
(s-s')} \tilde{B}_s \tilde{B}^\dg_{s'} - |a_1|^2 \rme^{-\rmi %\epsilon 
(s-s')} \tilde{B}^\dg_s \tilde{B}_{s'} \Big) \rho_\e \Big] \nonumber \\
=&- \frac{1}{\lambda^2}  \frac{\avg{\sigma_{(\theta,\phi)}}_\beta}{2}\sum_{nmpq} \Bigg( |a_0|^2  \bra{n_0}B\ketbra{p_1}{p_1} B^\dg \ketbra{m_0}{m_0} \rho_\e \ket{n_0} \int_0^{\lambda \tau} \rmd s \, \rme^{\rmi (1%\epsilon
+(E_n^0-E_p^1)/\lambda) s} \int_0^{\lambda \tau} \rmd s' \,  \rme^{-\rmi (1%\epsilon
+(E_m^0-E_p^1)/\lambda) s'} \nonumber \\
& \qquad\qquad\;\;- |a_1|^2  \bra{n_1} \rho_\e \ketbra{m_1}{m_1}B^\dg\ketbra{p_0}{p_0} B \ket{n_1} \int_0^{\lambda \tau} \rmd s \, \rme^{\rmi (1%\epsilon
-(E_n^1-E_p^0)/\lambda) s} \int_0^{\lambda \tau} \rmd s' \,  \rme^{-\rmi (1%\epsilon
-(E_m^1-E_p^0)/\lambda) s'} \Bigg) \nonumber \\
= &  - \frac{1}{\lambda^2 %\epsilon^2
} \frac{\avg{\sigma_{(\theta,\phi)}}_\beta}{2} \Bigg( |a_0|^2 \left(\avg{ B B^\dg}_\e + \avg{\rme^{\rmi A_0 \tau} B B^\dg \rme^{-\rmi A_0 \tau}}_\e - \rme^{\rmi \lambda %\epsilon 
\tau}\avg{\rme^{\rmi A_0 \tau} B \rme^{-\rmi A_1 \tau} B^\dg }_\e \right)  \nonumber \\
&\qquad\qquad\qquad\quad-|a_1|^2\left( \avg{B^\dg B}_\e + \avg{ \rme^{\rmi A_1 \tau} B^\dg B \rme^{-\rmi A_1 \tau}}_\e  -\rme^{\rmi \lambda %\epsilon 
\tau} \avg{ B^\dg \rme^{\rmi A_0 \tau} B \rme^{-\rmi A_1 \tau} }_\e\right)  \Bigg) +\mathcal{O}\left(\frac{1}{\lambda^3}\right), \label{eq:secondorder}
\end{align}
where again, the denominator has been expanded in powers of $1/\lambda$ after integrating. In the last line, we see the terms corresponding to $\xi^{(2)}_0(\lambda)$ and $\xi^{(2)}_1(\lambda)$ defined in Eqs.~\eqref{eq:xi21}~\&~\eqref{eq:xi22} respectively.

\section{Summary of obtainable properties of the system}
\label{sec:spectrum}
\begin{table}
\begin{tabular}{c|llr}
\hline 
 & System quantity & Fourier transform & Significance \\
\hline\hline
1 & $\avg{\rme^{-\rmi A_0 \tau} \rme^{\rmi A_1 \tau}}_\e$ &  $\sum_{nm} \bra{n_1}\rho_\e\ket{m_0}\braket{m_0}{n_1} \delta\big(\omega - (E_n^1-E_m^0)\big)$ & Equal to 1 when $A_0=A_1=H_\e$.\\
2 & $\avg{\rme^{\rmi A_0 \tau}B \, \rme^{-\rmi A_1 \tau}}_\e$ & $\sum_{nm} \bra{n_0}B\ketbra{m_1}{m_1}\rho_\e\ket{n_0} \delta\big(\omega - (E_n^0-E_m^1)\big)$ &   \\
3 & $\avg{\rme^{\rmi A_0 \tau}\rme^{-\rmi A_1 \tau} B^\dg}_\e$ & $ \sum_{nm} \braket{m_1}{n_0} \bra{n_0} B^\dg \rho_\e\ket{m_1} \delta\big(\omega - (E_n^0-E_m^1)\big)$ & \\
4 & $\avg{B^\dg \rme^{\rmi A_0 \tau}\rme^{-\rmi A_1 \tau}}_\e$ & $\sum_{nm} \braket{n_0}{m_1} \bra{m_1} \rho_\e B^\dg \ket{n_0} \delta\big(\omega - (E_n^0-E_m^1)\big)$ &  \\
5 & $\avg{\rme^{\rmi A_k \tau} B^\dg \rme^{-\rmi A_k \tau}}_\e$ & $\sum_{nm} \bra{n_k} B \ket{m_k}\bra{m_k}\rho_\e\ket{n_k} \delta\big(\omega - (E_n^k-E_m^k)\big)$ & Witness for initial system coherence.\\
6 & $\avg{\rme^{\rmi A_0 \tau}B \, \rme^{-\rmi A_1 \tau}B^\dg}_\e$ &  $\sum_{nm} \bra{n_0} B\ketbra{m_1}{m_1}B^\dg\rho_\e\ket{n_0} \delta\big(\omega - (E_n^0-E_m^1)\big)$ & Two-time correlation functions. \\
7 & $\avg{B^\dg\rme^{\rmi A_0 \tau} B \rme^{-\rmi A_1 \tau}}_\e$ & $\sum_{nm} \bra{n_0} B\ketbra{m_1}{m_1}\rho_\e B^\dg\ket{n_0} \delta\big(\omega - (E_n^0-E_m^1)\big)$ & \\
8 & $\avg{\rme^{\rmi A_0 \tau}BB^\dg\rme^{-\rmi A_0 \tau}}_\e$ & $\sum_{nm} \bra{n_0} B B^\dg \ketbra{m_0}{m_0}\rho_E\ket{n_0} \delta\big(\omega - (E_n^0-E_m^0)\big)$ & \\
9 & $\avg{\rme^{\rmi A_1 \tau}B^\dg B \, \rme^{-\rmi A_1 \tau}}_\e$ & $\sum_{nm} \bra{n_1} B^\dg B \ketbra{m_1}{m_1}\rho_E\ket{n_1} \delta\big(\omega - (E_n^1-E_m^1)\big)$ &  \\
\hline 
\end{tabular}
\caption{{\bf Properties of the system deducible from probability oscillations in $\lambda$}. Fourier distributions also presented. Here we have written the environment operators in their eigenbases as $A_k = \sum_n E_n^k \ket{n_k}\!\bra{n_k}$.
\label{tab:summaryfourier}}
\end{table}
Table~\ref{tab:summaryfourier} shows a summary of all the information one can obtain about the environment by measuring the amplitude, phase and displacement of probability oscillations in $\lambda$ as a function of  $\tau$. The functions have been Fourier transformed to give a distribution in energy; for a system with a discrete spectrum, each peak in the distribution will correspond to pairs of eigenstates of $A_0$ or $A_1$. In general, there will be multiple such pairs contributing to each peak, but for highly disordered environments (randomly distributed energy levels), each pair of levels will have its own peak. 

When no two energy differences in the system are the same, one effectively has access to each of the terms in the energy distributions listed in Table~\ref{tab:summaryfourier}. Furthermore, the whole spectrum of the operators $A_0$ and $A_1$ can be identified by comparing the different frequencies present in the distributions in row 5; triplets of energy levels can be identified from pairs of frequencies which sum to a third (e.g. $(E_n - E_m) + (E_m - E_p) = E_n-E_p$). By finding all such triplets which include the lowest and highest energy states (corresponding to the highest frequency present in the distribution), the entire spectrum can be reconstructed.

With information of the spectrum, one can pick out all the terms in, for example, the first row which come from the same energy level $E^1_n$. The sum of these terms gives the diagonal elements of $\rho_E$ in the basis of $A_0$, $\bra{n_0} \rho_\e \ket{n_0}$. Similarly, the diagonal elements in the basis of $A_1$ can also be found. Weighting the diagonal matrix elements by the corresponding energy eigenvalues allows the expectation values $\langle A_0 \rangle$ and $\langle A_1 \rangle$ to be calculated. Importantly, the sum of these gives the mean energy of the environment $\langle H_E \rangle = \langle A_0 \rangle + \langle A_1 \rangle$. However, it should be noted that this is only possible when there is coherence in the initial system state -- something that it is always possible to achieve by rotating the control basis, i.e., choosing different values of $\theta$ and $\phi$.

\section{Vibrational modes: derivation of probabilities}
\label{olf-details}
Here we outline the details of the derivation of the probabilities in subsection \ref{harmonic-oscillators}. We start with repeating the Hamiltonian of the molecular junction spin-boson model:
\begin{align}
 H=&\frac{1}{2}\lambda %\epsilon_{da}
 \left(\ketbra{d}{d}-\ketbra{a}{a}\right)+V(\ketbra{d}{a}+\ketbra{a}{d})
   %+(\gamma_d\ketbra{d}{d}+\gamma_a\ketbra{a}{a})\otimes(b_0+b_0^\dagger)\nonumber\\
   + \sum_k(\gamma_{dk}\ketbra{d}{d}+\gamma_{ak}\ketbra{a}{a})\otimes(b_k+b_k^\dagger)%+\omega_{\e}b_0^\dagger b_0
   +\sum_k\omega_kb_k^\dagger b_k,
\end{align}
with the relevant parameters explained in the core text -- note that, for convenience, we have set the zero of energy to be halfway between the donor and acceptor states of the probe. Next, we apply the polaron transformation $H\rightarrow e^S H e^{-S}$, generated by:
\begin{align}
    %S &= (u_d \ketbra{d}{d} - u_a \ketbra{a}{a}) \otimes (b_0^\dagger -b_0),\nonumber\\
    S&=\sum_k (u_{dk} \ketbra{d}{d} - u_{ak} \ketbra{a}{a}) \otimes (b_k^\dagger -b_k),
\end{align}
This transformation leaves operators $\ketbra{X}{X}_{X=d,a}$ unchanged but modifies the Hamiltonian (up to a constant energy shift) to:
\begin{align}
    &H=\frac{1}{2}\left(\lambda %\epsilon_{da}
    -E_r\right)\left(\ketbra{d}{d}-\ketbra{a}{a}\right)+H_{\se}, \quad \mbox{with} \\
    &H_{\se}=\ketbra{d}{a}\otimes B+\ketbra{a}{d}\otimes B^\dagger +\ketbra{d}{d}\otimes A_0+\ketbra{a}{a}\otimes A_1,
\end{align}
where 
$E_r=\sum_k\omega_k\left(u_{dk}^2-u_{ak}^2\right)$ and, written explicitly, the operators $A_{0,1}$ and $B$ relevant for our formalism are 
\begin{align}
    &A_0=A_1=%\omega_{\e}b_0^\dagger b_0+
    \sum_k\omega_kb_k^\dagger b_k,\\
    &B=V %D(u_d-u_a)
    \Pi_kD(u_{dk}-u_{ak}).
\end{align}
In the above, the displacement operator is defined as $D(\xi)=\exp\{\xi b^\dagger-\xi^*b\}$. A natural basis for preparations and measurements is the basis $\ketbra{X}{X}_{X=d,a}$. As a consequence, only the second-order terms play a role in the analysis of the probabilities $p_{\beta: \alpha}$. Since we have $B^\dg B = \mathbbm{1}$, the probabilities simplify to:
\begin{align}
    p_{\beta: \alpha}=& q_{\beta: \alpha} - \frac{\avg{\sigma_z}_\beta}{\lambda^2%\epsilon^2
    } \big(|a_0|^2-|a_1|^2\big) 
     \Re\left\{2V^2 - e^{i(\lambda%\epsilon
     -E_r)\tau} \avg{B^\dagger e^{iA_0\tau} B e^{-iA_0\tau}}\right\},
\end{align}
where $q_{\beta: \alpha}\in\{0,1\}$, $(|a_0|^2-|a_1|^2)=\pm 1$ and $\avg{\sigma_z}_\beta=\pm 1$, depending on the values of $\alpha$ and $\beta$. Restricting ourselves to transitions from $d$ to $a$ and substituting in the definition of $B$, we can further write:
\begin{align}
    p_{a: d} =& \frac{V^2}{2\lambda^2 %\epsilon^2
    } \times \big[ 2 - e^{i(\lambda%\epsilon
    -E_r)\tau+2 {\textstyle \sum_j} (u_{dj}-u_{aj})^2 \sin (\omega_j \tau)}  \avg{\Pi_jD\left((u_{dj}-u_{aj})(e^{i\omega_j\tau}-1)\right)}\big].
\end{align}

To analyze the constraints on the applicability of the method in this scenario, we first consider a transition element 
$|\langle m_1 m_2\cdots m_k \cdots|B|n_1 n_2\cdots n_k \cdots\rangle|$, where the $k$th mode initially has occupation number $n_k$ and ends up with occupation number $m_k$. This quantity must be much smaller than $\lambda%\epsilon
$ for states with non-negligible initial probability (and not too large for other states). Using the well known expression for the matrix elements of displacement operators~\cite{CahillGlauber1969}, we can write the constraint on $\lambda$ as
\begin{align}
\lambda %\epsilon
\gg V e^{-\frac{1}{2}\sum_k(u_{dk}-u_{ak})^2} \Pi_k \left(\frac{n_k!}{m_k!}\right)^{\frac{1}{2}} |u_{dk}-u_{ak}|^{m-n} L_n^{m-n}\left(|u_{dk}-u_{ak}|^2\right),
\end{align}
where $L_q^p(x)$ is a Laguerre polynomial. Unless there are particularly strongly coupled low frequency modes (corresponding to a sub-Ohmic spectral density, for which the polaron transformation is known to have problems), the satisfaction of this inequality principally depends on the value of $V$, which we have chosen to be small.
The second condition related to the validity of our expansion is given in Eq.~\eqref{eq:constraintsecond}. Taking advantage of the fact that in this example $B^\dagger=B^{-1}$ we can easily show that the relation between $\lambda$ and $\tau$ simplifies to: $\lambda%\epsilon
\gg\tau V^2$.

With the additional assumption on the form of system and the reservoir states being both thermal, we can evaluate the expectation value of the displacement operators to get the suppressing factor:
\begin{align}
   &\avg{\Pi_jD\left((u_{dj}-u_{aj})(e^{i\omega_j\tau}-1)\right)}=\exp\{-{\textstyle \sum_j} (u_{dj}-u_{aj})^2(1-\cos\omega_j\tau)\coth\left(\frac{\omega_j}{2k_B T}\right)\},
\end{align}
with $\beta_B$ being the inverse temperature Boltzmann factor. This allows us to rewrite the probabilities $p_{\beta: \alpha}$ in a much simpler form presented in the main text in Eq.~\eqref{olf-probability}.

\end{widetext}

\bibliography{ZenoOscillations}
\end{document}